\documentclass[preprint]{aastex}

\usepackage{graphicx}
\usepackage{array}
\usepackage{natbib}
\usepackage{amssymb}
\usepackage{lscape}

\begin{document}

\title{Finding $\eta$ Car Analogs in Nearby Galaxies Using \textit{Spitzer}: \\ 
II. Identification of An Emerging Class of Extragalactic Self-Obscured Stars}

\author{Rubab~Khan\altaffilmark{1},
C.~S.~Kochanek\altaffilmark{1,2},
K.~Z.~Stanek\altaffilmark{1,2},
Jill~Gerke\altaffilmark{1}
}

\altaffiltext{1}{Dept.\ of Astronomy, The Ohio State University, 140
W.\ 18th Ave., Columbus, OH 43210; khan, kstanek, ckochanek@astronomy.ohio-state.edu}

\altaffiltext{2}{Center for Cosmology and AstroParticle Physics, 
The Ohio State University, 191 W.\ Woodruff Ave., Columbus, OH 43210}

\shorttitle{Finding $\Eta$ Car}

\shortauthors{Khan et al.~2013}

\begin{abstract}
\label{sec:abstract}
Understanding the late-stage evolution of the most massive stars such as $\eta$\,Carinae 
is challenging because no true 
analogs of $\eta$\,Car have been clearly identified in the Milky Way or other galaxies. 
In \citet{ref:Khan_2013}, we utilized 
\textit{Spitzer} IRAC images of $7$ nearby ($\lesssim4$\,Mpc) galaxies to search for such analogs, 
and found $34$ candidates with flat or red mid-IR spectral energy 
distributions. Here, in Paper\,II, 
we present our characterization of these 
candidates using multi-wavelength data from the optical through the far-IR. 
Our search detected no true analogs of $\eta$\,Car, which implies an eruption rate that 
is a fraction $0.01\lesssim F \lesssim 0.19$ of the ccSN rate. This is 
roughly consistent with each $M_{ZAMS}~\gtrsim~70M_\odot$ star 
undergoing $1$ or $2$ outbursts in its lifetime. 
However, we do identify a significant population of 18 lower luminosity 
$\left(\log(L/L_\odot)\simeq5.5-6.0\right)$ dusty stars. Stars enter this phase at a 
rate that is fraction $0.09 \lesssim F \lesssim 0.55$ of the ccSN rate, 
and this is consistent 
with all $25~<~M_{ZAMS}~<~60M_\odot$ stars undergoing 
an obscured phase at most lasting a few thousand years 
once or twice. 
These phases constitute a negligible 
fraction of post-main sequence lifetimes of massive stars, which implies 
that these events are likely to be associated with special 
periods in the evolution of the stars. 
The mass of the obscuring material is of order 
$\sim M_\odot$, and we simply do not find 
enough heavily obscured stars for theses phases to represent more 
than a modest fraction ($\sim 10\%$ not $\sim 50\%$) 
of the total mass lost by these stars. 
In the long term, the sources that we identified 
will be prime candidates for detailed physical analysis with \textit{JWST}.
\end{abstract} 

\keywords{stars: evolution, mass-loss, winds, outflows
--- stars: individual: Eta Carinae
--- galaxies: individual (M33, M81, NGC247, NGC300, NGC2403, NGC6822, NGC7793)}
\maketitle

\section{Introduction}
\label{sec:introduction}

Despite being very rare, massive stars such as luminous blue variable (LBVs),
red super giants (RSGs), and Wolf-Rayet stars (WRs) play a pivotal role in
enriching the interstellar medium (ISM) through mass loss, and
they are an important source of heavier
elements contributing to the chemical enrichment of galaxies
\citep[e.g.,][]{ref:Maeder_1981}.
The deaths of these massive stars are associated with some of the highest energy 
phenomena in the universe such as core-collapse supernovae
\citep[ccSNe,][]{ref:Smartt_2009}, long-duration gamma-ray 
bursts \citep[e.g.,][]{ref:Stanek_2003}, 
neutrino bursts \citep[e.g.,][]{ref:Bionta_1987} and gravitational 
wave bursts \citep[e.g.,][]{ref:Ott_2009}. 
The physical mechanism, energetics and 
observed properties of these events depend on the structure and 
terminal mass of the evolved stars at core-collapse, which in turn are 
determined by stellar mass loss \citep[see, e.g., review by][]{ref:Smith_2014}. 
In addition, there is also evidence that some supernova (SN) progenitors undergo 
major mass ejection events shortly before exploding 
\citep[e.g.,][]{ref:GalYam_2007,ref:Smith_2008,ref:Ofek_2013a}, further altering 
the properties of the explosion and implying a connection between some eruptive 
mass-loss events and death. It is generally agreed that the effects of winds 
are metallicity dependent~\citep[e.g.,][]{ref:Meynet_1994,ref:Heger_2003} and the SNe 
requiring a very dense circumstellar medium 
\citep[e.g.,][]{ref:Schlegel_1990,ref:Filippenko_1997} predominantly 
occur in lower metallicity galaxies 
\citep[e.g.,][]{ref:Stoll_2011}. This strongly suggests that the 
nature and distribution of stars undergoing impulsive mass loss will also be 
metallicity dependent and a full understanding requires exploring galaxies 
beyond the Milky Way.

Understanding the evolution of massive (M$\gtrsim30 $\,M$_\odot$) stars is 
challenging even when mass loss is restricted to continuous winds 
\citep[e.g.,][]{ref:Fullerton_2006}. However, shorter, episodic eruptions, 
rather than steady winds, may be the dominant mass loss mechanism in the 
tumultuous evolutionary stages toward the end of the lives of the most massive 
stars \citep[e.g.,][]{ref:Humphreys_1984,ref:Smith_2006} as they undergo 
periods of photospheric instabilities leading to stellar transients 
($M_V\lesssim-13$) followed by rapid ($\dot{M}\gtrsim 10^{-4}\,M_\odot$year) 
mass-loss in the last stages of their evolution 
\citep[see][]{ref:Kochanek_2012a,ref:Smith_2014}. Deciphering the 
rate of these eruptions and their consequences is challenging because 
no true analog of $\eta$\,Car in mass, luminosity, energetics, mass lost and age has 
been found (see \citealp{ref:Smith_2011,ref:Kochanek_2012a}), 
and the associated transients are significantly fainter than supernova explosions 
and are easily missed. These phases are as difficult to model theoretically as 
they are to simulate computationally.

Dense winds tend to form dust,
although for hot stars the wind must be dense enough to form a pseudo-photosphere
in the wind~\citep{ref:Davidson_1987} that shields the dust formation region from the UV emission of the star
\citep{ref:Kochanek_2011c}. The star will then be heavily obscured by dust for an extended
period after the eruption (see, e.g., \citealp{ref:Humphreys_1994}).
The Great Eruption of $\eta$\,Car between 1840 and 1860 is the most 
studied case of a stellar outburst (see, e.g., \citealp{ref:Humphreys_2012}).
The $\sim10 M_{\odot}$ ejecta are now seen as a dusty nebula around
the star absorbing and then reradiating $\sim90\%$ of the light in the mid-IR.
This means that dusty ejecta are a powerful and long-lived signature of eruption.
The emission from these dusty envelopes peaks
in the mid-IR with a characteristic red color and a rising or flat
spectral energy distribution (SED) in the \textit{Spitzer} IRAC~\citep{ref:Fazio_2004} bands.

In the Galaxy, stars with resolved shells of dust emission are easily found at
24\,$\micron$~\citep{ref:Wachter_2010,ref:Gvaramadze_2010}.
The advantage of the 24\,$\micron$ band is that it can be used to identify dusty ejecta
up to $10^3 - 10^4$\,years after formation. A minority of these objects are very luminous stars
(L\,$\gtrsim10^{5.5}\,$\,L$_\odot$) with massive ($\sim0.1-10\,$\,M$_\odot$) shells (see summaries by
\citealp{ref:Humphreys_1994,ref:Humphreys_1999,ref:Smith_2006,ref:Smith_2009,ref:Vink_2009}).
These include AG\,Car~\citep{ref:Voors_2000},
the Pistol Star~\citep{ref:Figer_1999}, G79.29$+$0.46~\citep{ref:Higgs_1994},
Wray\,17$-$96~\citep{ref:Egan_2002}, and IRAS\,18576$+$0341~\citep{ref:Ueta_2001}.
These systems are significantly older ($10^3-10^4$\,years) than $\eta$\,Car, which makes it difficult
to use the ejecta to probe the rate or mechanism of mass-loss.
Still, the abundance of Galactic shells implies that the rate of
$\eta$\,Car-like eruptions is on the order of a modest fraction of the ccSN rate~\citep{ref:Kochanek_2011c}.
Their emission peaks
in the shorter IRAC bands when they are relatively young ($\sim10-100$\,years)
because the dust becomes cooler and the emission shifts to longer wavelengths
 as the ejected material expands \citep{ref:Kochanek_2012a}.
It is difficult to quantify searches for such objects in our Galaxy because it is
hard to determine the distances and the survey
volume because we have to look through the crowded and dusty disk of the Galaxy.
Surveys of nearby galaxies are both better defined and can be used to build larger samples
of younger systems whose evolution can be studied to better understand the mechanism.
We previously demonstrated in \citet{ref:Khan_2010,ref:Khan_2011} that it is 
possible to identify post-eruptive massive stars in galaxies beyond the Local 
Group using the mid-IR excess created by warm circumstellar dust despite 
the crowding problems created by the limited spatial resolution of \textit{Spitzer} 
at greater distances. 

In \citet{ref:Khan_2013} (``Paper\,I'' hereafter) we used archival \textit{Spitzer} IRAC 
images of seven $\lesssim4$\,Mpc galaxies (closest to farthest: NGC\,6822, M\,33, NGC\,300, 
NGC\,2403, M\,81, NGC\,0247, NGC\,7793) in a pilot study to search for 
extragalactic analogs of $\eta$\,Car. We found 34 candidates with flat or 
rising mid-IR spectral energy distributions (SEDs) and total mid-IR luminosity 
$L_{mIR}\gtrsim 5\times10^5 L_\odot$. 
Here, in Paper\,II, we characterize 
these sources and quantify the rate of episodic
mass loss from massive stars in the last stages of evolution.
First, we construct extended optical through far-IR SEDs 
using archival HST, 2MASS, and \textit{Herschel} data as well as 
ground based data (Section\,\ref{sec:data}). Then, we classify the sources as either stellar or non-stellar 
based on properties of the extended SEDs and model the SEDs 
to infer the properties of the underlying star and the obscuring circumstellar 
medium (Section\,\ref{sec:analysis}). Next, we relate these properties to the observed  
ccSN rate of the targeted galaxies to quantify the rate of episodic 
mass loss in the last stages of massive star evolution
(Section\,\ref{sec:discussion3}). Finally, we
consider the implications of our
findings for theories and observations of massive star evolution and their fates 
(Section\,\ref{sec:conclusions}).

\section{Additional Wavelength Coverage}
\label{sec:data}

\begin{figure}[p]
\begin{center}
\includegraphics[angle=0,width=140mm]{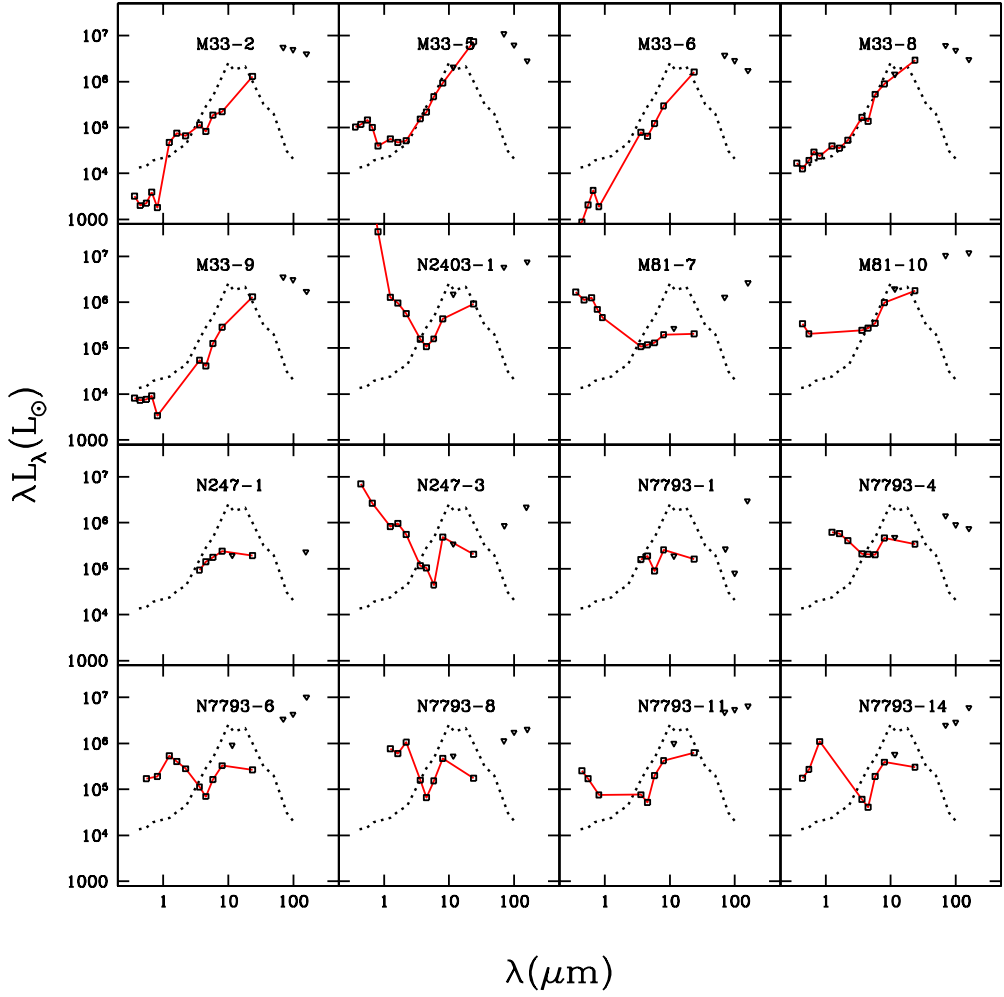}
\end{center}
\caption{The SEDs of the 16 candidates that we concluded are not stars 
(points and solid lines) as compared to $\eta$\,Car (dotted line).}
\label{fig:reject}
\end{figure}

In this Section, we describe the details of how we obtained the photometric 
measurements at various wavelengths to determine the properties of the 
candidates from Paper\,I. 
The optical through far-IR photometry are reported in 
Table\,\ref{tab:photo}, and the extended SEDs are shown in Figures 
\ref{fig:accept} and \ref{fig:reject}.

We utilized VizieR\footnote{http://vizier.u-strasbg.fr/} 
\citep{ref:Vizier} to search for other observations of the candidates, in 
particular for WISE (\citealt{ref:Wright_2010}, $12\micron$), 2MASS 
(\citealt{ref:Cutri_2003}, $JHK_s$), SDSS (\citealt{ref:Adelman_2009}, $ugriz$) 
and X-ray detections. For M\,33, we used the $UBVRI$ images from the 
\citet{ref:Massey_2006} optical survey, and archival HST images of 
NGC\,300, NGC\,2403, M\,81, NGC\,247 and NGC\,7793. Finally, we used 
\textit{Herschel} PACS data to supplement the \textit{Spitzer} measurements. 

For the \textit{Spitzer} IRAC 3.6, 4.5, 5.8 and $8\micron$ as well as MIPS 
\citep{ref:Reike_2004} 24, 70, and $160\micron$ data, we use the 
measurements reported in Paper\,I. For M\,33, our measurements were 
based on IRAC data from \citet{ref:McQuinn_2007} 
and MIPS data from the 
\textit{Spitzer Heritage Archive}\footnote{http://sha.ipac.caltech.edu/applications/Spitzer/SHA/}.
Data from the LVL survey \citep{ref:Dale_2009} were used for NGC\,300 and 
NGC\,247, and data from the SINGS survey \citep{ref:Kennicutt_2003} for 
NGC\,6822, NGC\,2403, and M\,81. 

\begin{figure}[t]
\begin{center}
\includegraphics[angle=0,width=150mm]{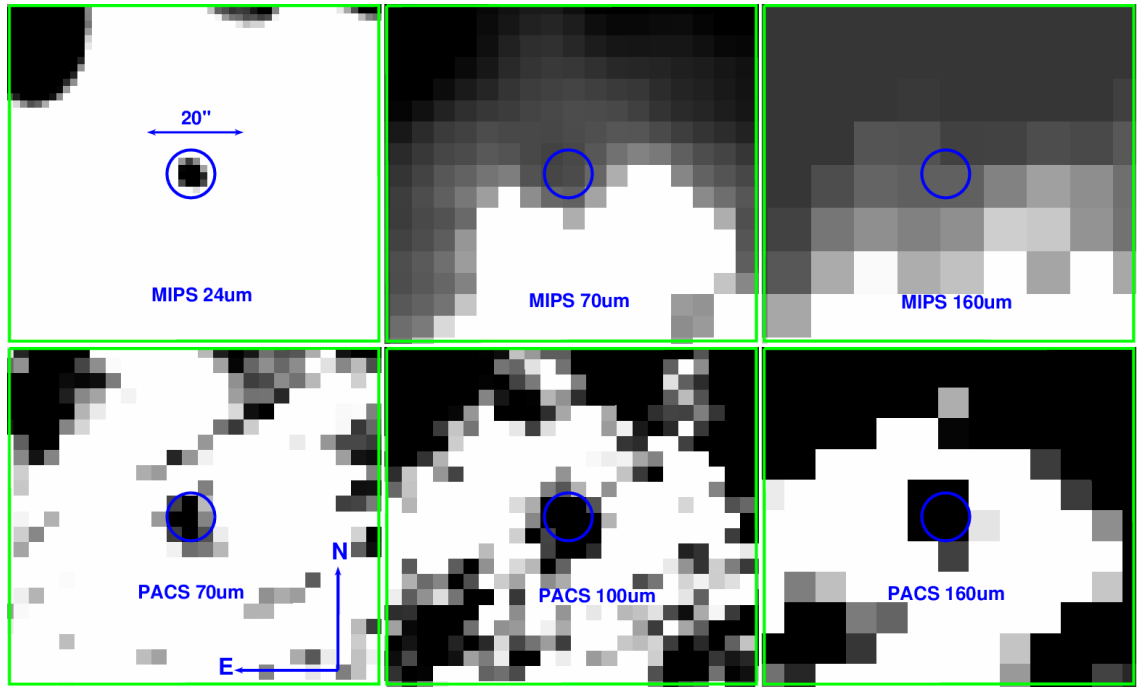}
\end{center}
\caption{The \textit{Spitzer} MIPS 24, 70 and $160\,\micron$ (top row) 
and \textit{Herschel} PACS 70, 100 and $160\,\micron$ (bottom row) images 
of the region around the object N7793-9. The higher resolution of the 
PACS images helps us set tighter limits on the far-IR emission from 
the candidates.}
\label{fig:mips_pacs}
\end{figure}

We used the \textit{Herschel} PACS \citep{ref:Poglitsch_2010} 70, 100, and 
$160\micron$ images available from the public 
\textit{Herschel Science Archive}\footnote{http://herschel.esac.esa.int/Science\_Archive.shtml}.
Although both MIPS and PACS cover the same far-IR wavelength range 
($70-160\micron$), Herschel has significantly higher resolution 
(see Figure\,\ref{fig:mips_pacs}). All three PACS 
band data were available for M\,33 and NGC\,7793, 70 and 160$\micron$ data 
were available for NGC\,2403 and M\,81, and 100 and 160$\micron$ data were 
available for NGC\,300. There are no publicly available PACS images of the 
candidates in NGC\,247. We used aperture photometry (IRAF\footnote{IRAF is 
distributed by the National Optical Astronomy Observatory, which is operated by 
the Association of Universities for Research in Astronomy (AURA) under 
cooperative agreement with the National Science Foundation.} ApPhot/Phot) with 
the extraction apertures and aperture corrections from \citet{ref:Balog_2013} 
and given in Table\,\ref{tab:pacs}. As with our treatment of the MIPS 70 and 160$\micron$ 
measurements in Paper\,I, we treat the measurements obtained in the PACS bands 
as upper limits because the spatial resolution of these bands requires 
increasingly large apertures at longer wavelengths. For similar 
reasons, we also treat the WISE $12\micron$ fluxes, where available, as upper 
limits.
 
For the optical photometry of the candidates in M\,33, we used the Local Group 
Galaxies Survey $UBVRI$ images \citep{ref:Massey_2006}. First we verified that the coordinates
match with the IRAC images to within few$\times0\farcs1$ and then used $1\farcs0$ 
radius extraction apertures centered on the IRAC source locations. We 
transformed the aperture fluxes to Vega-calibrated magnitudes using zero point 
offsets determined from the difference between our aperture magnitudes and 
calibrated magnitudes for bright stars in the \cite{ref:Massey_2006} catalog 
of M\,33. 

For the candidates in NGC\,300, M\,81, NGC\,2403, and NGC\,247, we 
searched the ACS Nearby Galaxy Survey 
\citep[ANGST,][]{ref:Dalcanton_2009} $B$, $V$ and (where available) $I$ band 
point source catalogs derived 
using DOLPHOT \citep{ref:Dolphin_2000}.
We verified that the IRAC and HST astrometry of the NGC\,300, NGC\,2403 and NGC\,247 images agree
within (mostly) $\lesssim0\farcs1$ to (in a few cases) $0\farcs3$. 
We corrected the astrometry of the M\,81 HST images using the LBT images described later in this section 
to achieve similar astrometric accuracy. 
We also used the HST $I$-band
photometry of M\,81 from HST program GO-10250 (P.I. J.\,Huchra).
We retrieved all publicly available archival HST images of NGC\,7793 overlapping 
the IRAC source locations along with the associated photometry tables from the Hubble Legacy 
Archive\footnote{http://hla.stsci.edu/}. The HST and \textit{Spitzer}
images have a significant (few$\times 1\farcs0$) astrometric mis-match, and there are too few
reference stars in the HST images to adequately improve the astrometry. Therefore, we utilized the
IRAF GEOXYMAP and GEOXYTRAN tasks to locally match the overlapping HST and \textit{Spitzer}
images of NGC\,7793 within uncertainties of $0\farcs1 \sim 0\farcs3$.

\begin{figure}[p]
\begin{center}
\includegraphics[angle=0,width=165mm]{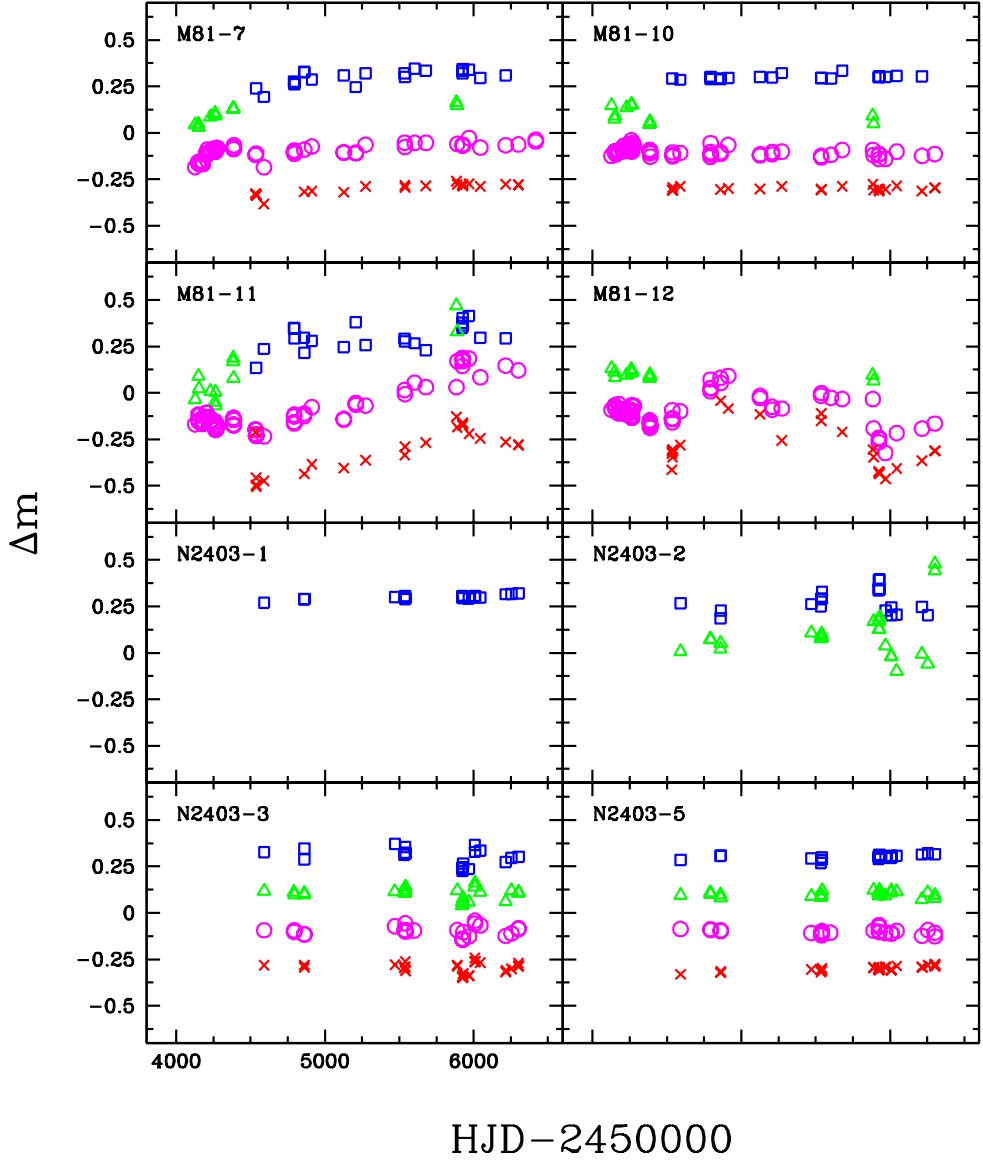}
\end{center}
\caption{The differential light curves of some of 
the candidates in M\,81 and NGC\,2403 obtained from 
the Large Binocular Telescope. 
The data spans the period from March\,2008 to 
January\,2013. The $U$ (squares), $B$ (triangles), 
$V$ (circles), $R$ (crosses) differential magnitudes
are offset by $+0.3, +0.1, -0.1, -0.3$\,mag for clarity.}
\label{fig:lc}
\end{figure}

We have variability data for the galaxies M\,81 and NGC\,2403 
from a Large Binocular Telescope survey in the $U B V R$ bands 
that is searching for failed supernovae 
\citep{ref:Kochanek_2008}, and studying supernova progenitors 
and impostors \citep{ref:Szczygiel_2012}, and 
Cepheid variables \citep{ref:Gerke_2011}. We analyzed 27 epochs of data for 
M\,81 and 28 epochs of data for NGC\,2403, spanning a 5\,year period. 
The images were analyzed with the ISIS image subtraction 
package \citep{ref:Alard_1998,ref:Alard_2000} to produce light 
curves (see Figure\,\ref{fig:lc}). 

\section{Characterizing the Candidates}
\label{sec:analysis}

In this section, we first discuss how we classify the candidates
based on their SEDs. Next, we describe the non-stellar and stellar 
sources. Finally, we model the SEDs of the stellar sources 
to determine their physical properties. Figure\,\ref{fig:accept} 
shows the SEDs of the stellar sources with the best fit SED models 
over plotted and
Figure\,\ref{fig:reject} shows the SEDs of the non-stellar sources.

\subsection{Source Classification}

We classify the candidates either as
stellar or non-stellar based on their photometric properties.
We focus on identifying two tell-tale signatures of the SED of 
a luminous star obscured by warm circumstellar dust --- 
low optical fluxes or flux-limits 
compared to the mid-IR luminosities and signs of the SEDs turning 
over between $8\,\micron$ and $24\,\micron$. 
Towards longer 
wavelengths, emission from warm circumstellar dust should peak 
between the IRAC $8\,\micron$ and MIPS $24\,\micron$ bands.
It is almost impossible for mass lost from a single star to 
to both have a significant optical depth and a dust temperature 
cold enough to peak at wavelengths longer than $\sim24\,\micron$.
Such systems are almost certainly star clusters with significant 
amounts of cold dust.
Therefore, any SED that appears 
to have a steep slope between $8$ and $24\,\micron$ is considered 
to be a likely cluster, rather than a single dust obscured star.
Frequently, these sources are also too luminous to be a single 
star. At the shorter wavelengths, we 
expect a dusty star to have relatively lower luminosity 
compared to its mid-IR luminosity and redder optical 
colors. 

We examine the HST $B-V$/$V$ and the $V-I$/$V$ color magnitude diagrams
(CMDs) for each source for which HST data is available. 
The presence of a very red optical 
counterpart or the absence of a luminous star supports the 
existence of significant dust obscuration. On the other hand, the presence of a blue 
or bright optical counterpart makes it likely that the 
source is a star cluster, a background galaxy/AGN, or a foreground star. 
We first search
for bright and/or red optical sources
within the $0\farcs3$ matching radius that can be the obvious counterpart of the bright and
red IRAC source. Next, if multiple bright and/or red optical matches are found,
we identify the best astrometric match to the IRAC location. Finally, if no
reasonable match is found, we adopt the flux of the brightest of the
nearby sources as a conservative upper limit on the optical luminosity of the
candidate. 

\begin{figure}
\begin{center}
\includegraphics[angle=0,width=130mm]{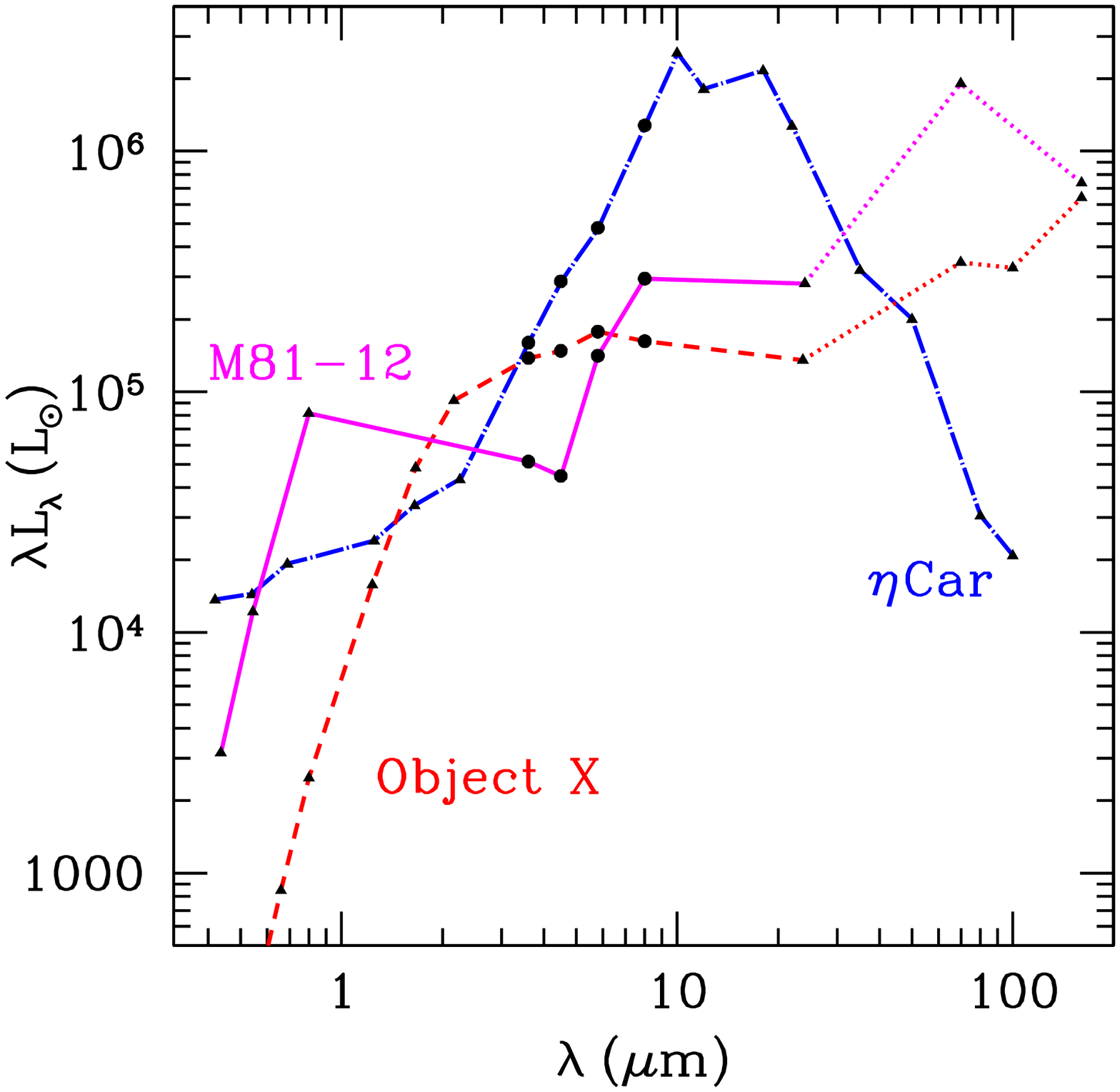}
\end{center}
\caption{The spectral energy distributions (SEDs) of the dust-obscured massive 
star $\eta$\,Car (dash-dot line, \citealp[e.g.,][]{ref:Robinson_1973,ref:Humphreys_1994}), 
``Object\,X'' in M\,33 \citep[dashed line;][]{ref:Khan_2011}, and 
an obscured star in M\,81 that we identify in this paper 
(M\,81-12, solid line). All these stars have SEDs that 
are flat or rising in the \textit{Spitzer} IRAC 3.6, 4.5, 5.8 and $8.0\,\micron$ 
bands (marked here by solid circles). 
The three shortest wavelength data-points of the M\,81-12 SED are from HST $BVI$ images.
The $24\,\micron$ measurements of both Object\,X and M\,81-12 are from 
\textit{Spitzer} MIPS while the dotted segments of their SEDs show the 
\textit{Herschel} PACS 70, 100, and $160\,\micron$ upper limits.}
\label{fig:Eta_sed}
\end{figure}

\begin{figure}
\begin{center}
\includegraphics[angle=0,width=150mm]{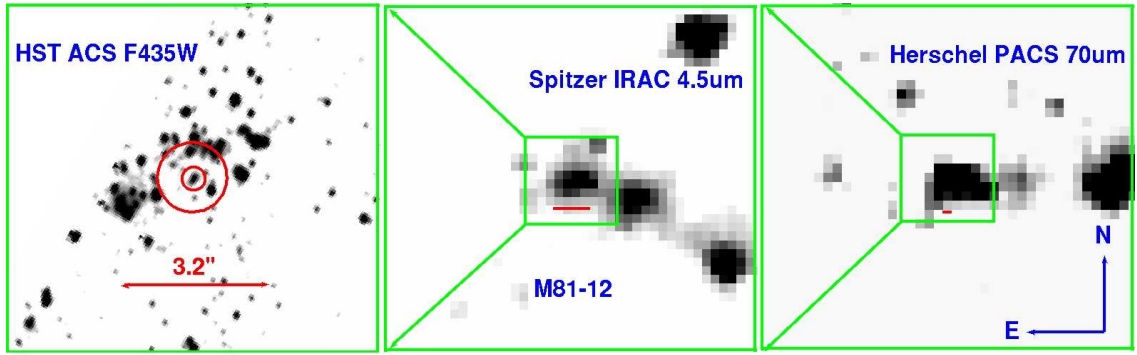}
\end{center}
\caption{The \textit{Hubble}, \textit{Spitzer}, and 
\textit{Herschel} images of the region around M\,81-12. 
In the left panel, the radii of the circles are $0\farcs25$ 
(5 ACS pixels) and $1\farcs43$ (IRAC 
$4.5\,\micron$ PSF FWHM), and the source at the position of the 
smaller circle in the left panel is the brightest red point 
source on the CMD (Figure\,\ref{fig:cmd}, \textit{left} panel). 
The red line in each panel is the size of a 
PACS pixel ($3\farcs2$).}
\label{fig:m81-12_all}
\end{figure}

\begin{figure}
\begin{center}
\includegraphics[angle=0,width=130mm]{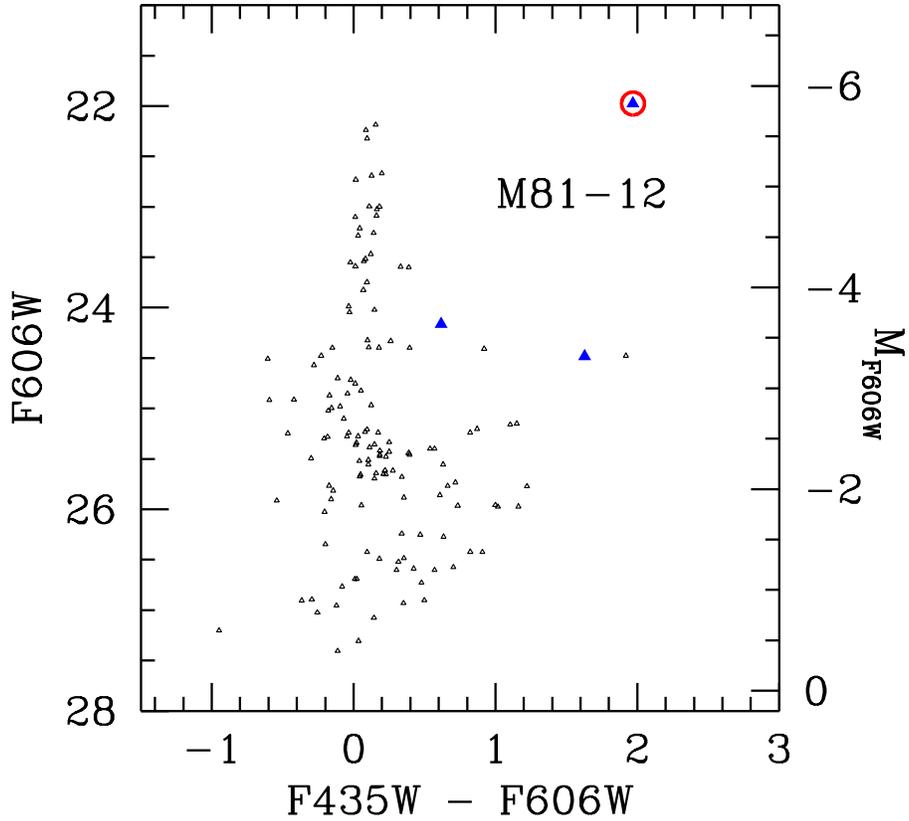}
\end{center}
\caption{The $F606W$ ($V$) vs. $F435W-F606W$ ($B-V$) color magnitude diagram (CMD) for all HST point 
sources around M\,81-12. The three large solid triangles denote sources located with the 
$0\farcs3$ matching radius. The small open triangles show all other sources within 
a larger $2\farcs0$ radius to emphasize the absence 
of any other remarkable sources nearby.
The circle marks the source at the position of the 
smaller circle in the left panel of Figure\,\ref{fig:m81-12_all},
which is the brightest red point source on the CMD. 
The excellent ($<0\farcs1$) astrometric 
match and the prior that very red sources are rare 
confirms that this source is the optical counterpart of the 
mid-IR bright red \textit{Spitzer} source.}
\label{fig:cmd}
\end{figure}

To demonstrate these, we discuss the case of M\,81-12 in detail.
M\,81-12 has a steeply rising optical and mid-IR SED
(Figure\,\ref{fig:Eta_sed}) with two distinct peaks --- one in the near-IR,
between the $R$-band and 3.6\,\micron, and another in the mid-IR between
8 and 24\,\micron. 
Figure\,\ref{fig:cmd} shows the HST optical CMD for sources
near the location of M\,81-12. Besides the sources within the $0\farcs3$
matching radius, it also shows all sources within $0\farcs3 - 2\farcs0$ of the candidate
using a different symbol to emphasize the absence of any other unusual
nearby sources.
We detect a very red ($B=23.95$, $V=21.98$, $I=19.07$,
$B-V\simeq2$, $V-I\simeq2.9$) HST counterpart with an excellent astrometric
match ($<0\farcs1$, Figure\,\ref{fig:m81-12_all}) to the IRAC position.
This source is the brightest,
red HST point source within $2\farcs0$ of the IRAC location
(Figure\,\ref{fig:cmd}) and so we define it to be the counterpart used in the SED. 
The LBT $V$ and $R$ band light curves show
a variable source with the correlated irregular variability ($\sim$0.4\,mag,
Figure\,\ref{fig:lc}) typical of many evolved massive stars \citep[e.g.,][]{ref:Kourniotis_2013}. Based on the 
SED shape and the unambiguous detection of a red, variable optical 
counterpart, we conclude that M\,81-12 is a massive, dust-obscured, single star.

In addition to Object\,X (M\,33-1), we identified
17 additional dust obscured stars and
classified 16 others as non-stellar. We left one source (N\,7793-12)
unclassified due to a lack of sufficient optical data 
(it falls on an HST/ACS chip gap). It could well be a dusty star, but we 
do not discuss it further.

\subsection{The 18 Stars and 16 Non-stellar Sources}
\label{sec:stars}

We identify 18 (including Object\,X/M\,33-1) sources as dusty stars. Of these, four
are in M\,33 (1, 3, 4, 7), one is in NGC\,300 (N\,300-1), four are in NGC\,2403 
(2, 3, 4, 5), five are in M\,81 (5, 6, 11, 12, 14), and four are in NGC\,7793 
(3, 9, 10, 13). None are in NGC\,6822 (a low-mass, low SFR galaxy) or 
NGC\,247 (all three candidates turned out to be non-stellar).
These stars have low optical fluxes or flux limits 
and their SEDs turn over between $8\,\micron$ and $24\,\micron$.
Moreover, M\,81-11, M\,81-12 and N\,2403-2 
are detected as optically variable sources in the LBT monitoring data. N\,2403-3 is a saturated source
in the HST images, and we use the LBT flux measurements as upper limits on its 
optical flux. N\,2403-3 and N\,2403-5 are not variable in the LBT data.
M\,81-5 is 0\farcs56 from a variable X-ray source
with maximum luminosity of $2\times10^{38}$\,ergs\,s$^{-1}$ \citep{ref:Liu_2011},
which is consistent with the source being an X-ray binary~\citep{ref:Remillard_2006}.
N\,7793-3 is also a X-ray source \citep{ref:Liu_2011}, with a maximum X-ray luminosity 
of $3.9\times10^{37}$erg$s^{-1}$ and is classified as an HMXB by \cite{ref:Mineo_2012}. 

There are 16 candidates whose SEDs indicate that they are not self-obscured stars.
Five sources in M\,33 (2, 5, 6, 8, 9) have SEDs that nearly monotonically rise 
from the optical to $24\,\micron$, unambiguously indicating the presence of cold 
dust associated with star clusters. As
we discussed in Paper\,I, it is unlikely for an ultra-compact
star cluster to host both evolved massive stars and significant
amounts of intra-cluster dust. 
Eight sources cannot be dust obscured stars given their 
very high optical luminosities: N\,2403-1 (likely a foreground star),
M\,81-7, M\,81-10, N\,247-3 (likely a foreground star, optical magnitudes 
from \citealp{ref:STSci_2001}), N\,7793-4,
N\,7793-8, N\,7793-11, and N\,7793-14. 
The three observed with the LBT, M\,81-7, M\,81-10 and N\,2403-1, are not variable.
We consider three more sources as most likely 
non-stellar due to reasons that are unique in each case~---
\begin{itemize}
\item{N\,247-1} is located far from the plane of its edge-on host
and is unlikely to be associated with the host.
\item{N\,7793-1} is located at the edge of its host galaxy
and the PACS far-IR
flux limits are significantly lower than those of the sources 
that we classified as obscured stars, indicating an absence of 
the diffuse emission commonly associated with star forming regions.
\item{N\,7793-6} has an SED that can conceivably be produced by a hot star 
with significant circumstellar material, although the near-IR peak seems 
too narrow. However, a close inspection of the 
HST image shows that this source is in a dense star-forming region with 
significant diffuse light indicating the presence of intra-cluster dust.
None of the sources in the HST image are a good astrometric match to the 
IRAC location.
It is more likely, in this case, that warm intercluster dust is producing 
the mid-IR flux excess. The optical fluxes adopted here are those of the 
most luminous HST source within a larger matching radius of $0\farcs5$.
\end{itemize}

In Paper\,I, we anticipated that further analysis would show that most, if not all, of the 
candidates are in fact non-stellar sources. Based on the expected surface density 
of extragalactic contaminants, of the 46 initial candidates
we estimated that all but $6\pm6$ are background galaxies/AGN
with 11 already being identified as such. Here we find that 
$18$ (including Object\,X) of the candidates are dusty massive stars and
very few of the other sources are background galaxies. We do not 
presently have an explanation for the fewer than expected background 
sources in the targeted fields.

\subsection{SED Modeling}

\begin{figure}
\begin{center}
\includegraphics[angle=0,width=130mm]{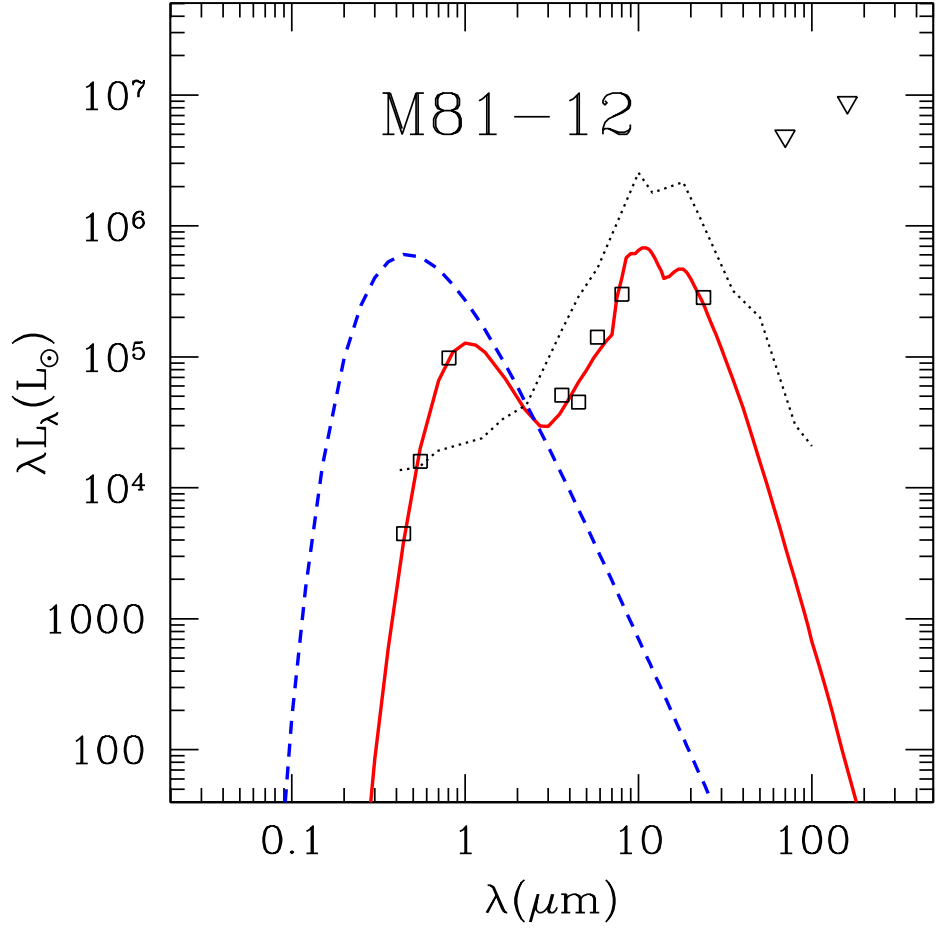}
\end{center}
\caption{The best fit model (solid line)
of the observed SED (squares and triangles, the latter show 
flux upper limits) of M\,81-12
and the SED of the 
underlying, unobscured star (dashed line),
as compared to 
$\eta$\,Car (dotted line).
The best fit is for a 
$L_*\simeq 10^{5.9}\,L_\odot$, $T_*\simeq7900\,K$ star
obscured by $\tau_V\simeq8$, $T_d\simeq530\,K$ silicate dust 
shell at $R_{in}=10^{16.1}$\,cm. }
\label{fig:Dusty2}
\end{figure}

\begin{figure}[p]
\begin{center}
\includegraphics[angle=0,width=165mm]{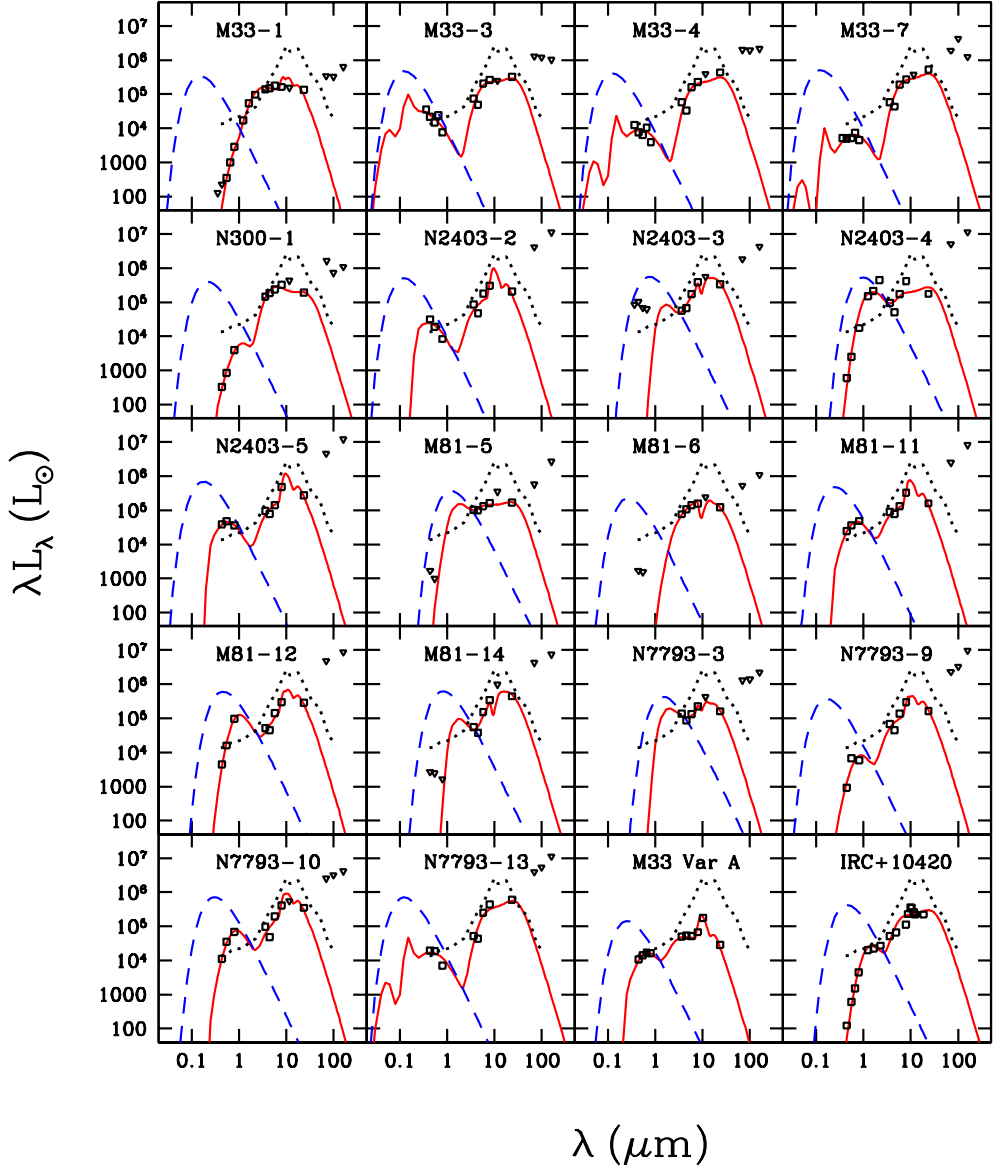}
\end{center}
\caption{Same as Figure\,\ref{fig:Dusty2}, but showing all the obscured 
stars that we identified as compared to M\,33\,Var\,A, $IRC+10420$, and 
$\eta$\,Car. The solid line shows the best fit model of the observed SED, 
and the dashed line shows the SED of the 
underlying, unobscured star. M\,33\,Var\,A and $IRC+10420$ are shown on 
separate panel while $\eta$\,Car is shown on every panel (dotted line).}
\label{fig:accept}
\end{figure}

We fit the SEDs of the 18 self-obscured stars using DUSTY 
\citep{ref:Ivezic_1997,ref:Ivezic_1999,ref:Elitzur_2001} to model radiation 
transfer through a spherical dusty medium surrounding a star and
Figure\,\ref{fig:accept} shows the best fit models.
We estimate the properties of a black-body source obscured by a surrounding dusty shell 
that would produce the best fit to the observed SED (see 
Figure\,\ref{fig:Dusty2} for an example). We considered models with either 
graphitic or silicate \citep{ref:Draine_1984} dust. We distributed the 
dust in a shell with a $\rho \propto 1/r^2$ density distribution. 
The models are defined by the stellar luminosity ($L_*$), 
stellar temperature ($T_*$), the total (absorption plus scattering) 
$V$-band optical depth ($\tau_V$), the dust 
temperature at the inner edge of the dust distribution ($T_d$), and the shell 
thickness $\zeta=R_{out}/R_{in}$.
The exact value of $\zeta$ has little effect on the results, and
after a series of experiments with $1<\zeta<10$, we fixed $\zeta=4$ for
the final results.
We embedded DUSTY inside a Markov 
Chain Monte Carlo (MCMC) driver to fit each SED by varying 
$T_*$, $\tau_V$, and $T_d$. We limit $T_*$ to a maximum value of 30,000\,K 
to exclude unrealistic temperature regimes. 

The parameters of the best fit model determine the radius of the inner edge of
the dust distribution ($R_{in}$).
The mass of the shell is 
\begin{equation}
\label{eqn:mejecta}
M_e = \frac{4 \pi R_{in}^2 \tau_V}{\kappa_V}
\end{equation}
where we simply scale the mass for a $V$ band dust opacity of 
$\kappa_V=100\,\kappa_{100}$~cm$^2~$g$^{-1}$ and the result can be 
rescaled for other choices as $M_e \propto \kappa_V^{-1}$.
Despite using a finite width shell, we focus on $R_{in}$ because it is 
well-constrained while $R_{out}$ (or $\zeta$) is not. We can also 
estimate an age for the shell as 
\begin{equation}
\label{eqn:age}
t_e = \frac{R_{in}}{v_e}
\end{equation}
where we scale the results to $v_e=100\,v_{e100}$\,km\,s$^{-1}$. 

For a comparison sample, we followed the same procedures 
for the SEDs of three well-studied dust obscured stars: 
$\eta$\,Car \citep{ref:Humphreys_1994}; 
the Galactic OH/IR star IRC+10420 \citep{ref:Jones_1993,ref:Humphreys_1997,ref:Tiffany_2010}; 
and M\,33's Variable\,A, which had a brief period of high mass loss leading to dust 
obscuration over the last $\sim50$\,years 
\citep{ref:Hubble_1953,ref:Humphreys_1987,ref:Humphreys_2006}. We use the same SEDs for these stars as in 
\citet{ref:Khan_2013}. 
In Table\,\ref{tab:mcmc}, we report $\chi ^ 2$, $\tau_V$, $T_d$, $T_*$, $R_{in}$, $L_*$, 
$M_e$ (Equation\,\ref{eqn:mejecta}), and $t_e$ (Equation\,\ref{eqn:age})
for the best fit models for these three sources as well as the newly 
identified stars. The stellar 
luminosities required for both dust types are mutually consistent
because the optically thick dust shell acts as a calorimeter. 
However, because the stars are heavily obscured and we have limited 
optical/near-IR SEDs, the stellar temperatures generally are not well constrained.
In some cases, for different dust types, 
equally good models can be obtained for either a hot ($>25000\,K$, such as a LBV in quiescence) or a 
relatively cooler ($<10000\,K$, such as a LBV in outburst) star. 
Indeed, for many of our 18 sources, the best fit 
is near the fixed upper limit of $T_* = 30000\,K$. 
To address this issue, we also tabulated the models on a grid of three fixed 
stellar temperatures, $T_* = 5000\,K, 7500\,K, 20000\,K$, for each dust type. 
The resulting best fit parameters are reported in Tables \ref{tab:graphitic} 
and \ref{tab:silicate}.

Figure\,\ref{fig:HR_like} shows the integrated luminosities 
of the newly identified self-obscured stars described in Section\,\ref{sec:stars}
as a function of $M_e$ 
for the best fit graphitic models of each source.
Object\,X, IRC$+10420$, M\,33\,Var\,A, and $\eta$\,Car are shown for comparison.
Figure\,\ref{fig:Dusty} shows the same quantities, but for various dust 
models and temperature assumptions.
It is apparent from Figure\,\ref{fig:Dusty} and Tables 
\ref{tab:mcmc}, \ref{tab:graphitic} and \ref{tab:silicate} that the integrated luminosity 
and ejecta mass estimates are robust to these uncertainties.
The exceptions are N\,2403-4 and N\,7793-3.
Without any optical or near-IR data, many of the models of N\,7793-3
are unstable so we simply drop it.
The only models having a luminosity in significant excess of 
$10^6\,L_\odot$ are some of the fixed temperature models of N\,2403-4.
These models have a poor goodness of fit and can be ignored.

\begin{figure}[p]
\begin{center}
\includegraphics[angle=0,width=130mm]{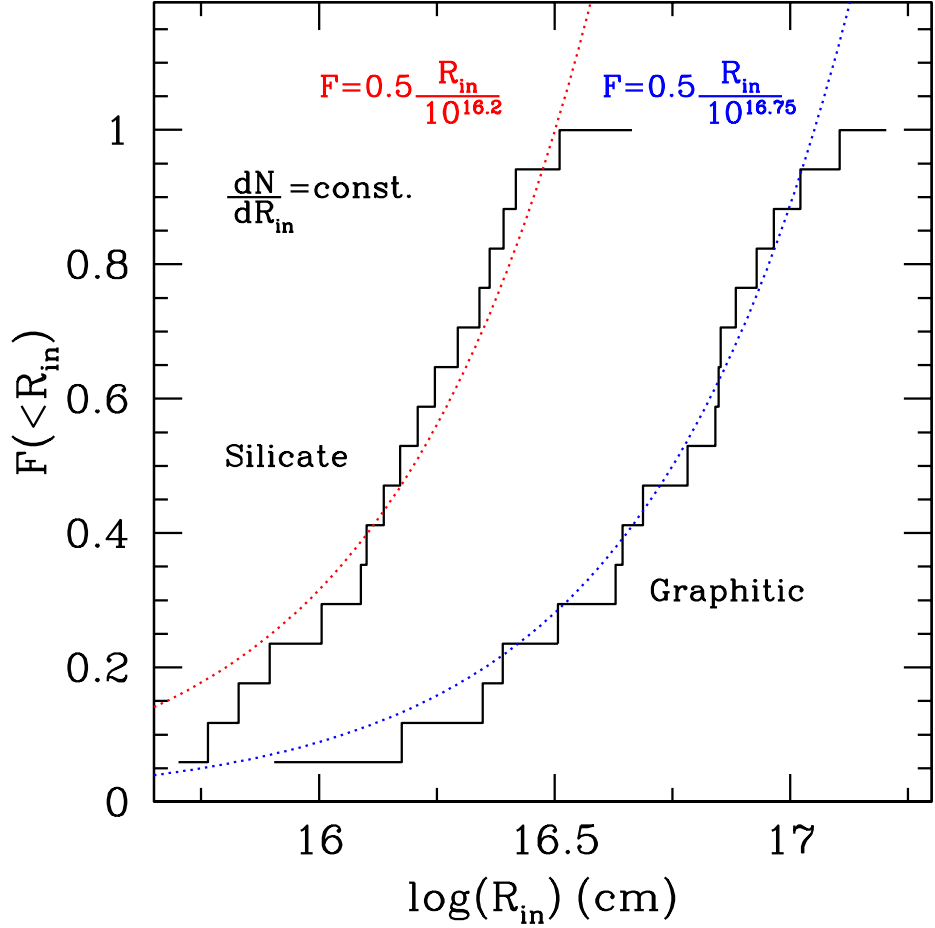}
\end{center}
\caption{Cumulative histograms of the dust shell radius $R_{in}$ for the newly identified stars 
excluding N\,7793-3. The dotted lines, normalized to the point where 
$F(<R_{in})=0.5$, shows the distribution expected for shells in 
uniform expansion observed at a random time.}
\label{fig:complete}
\end{figure}

\begin{figure}[p]
\begin{center}
\includegraphics[angle=0,width=130mm]{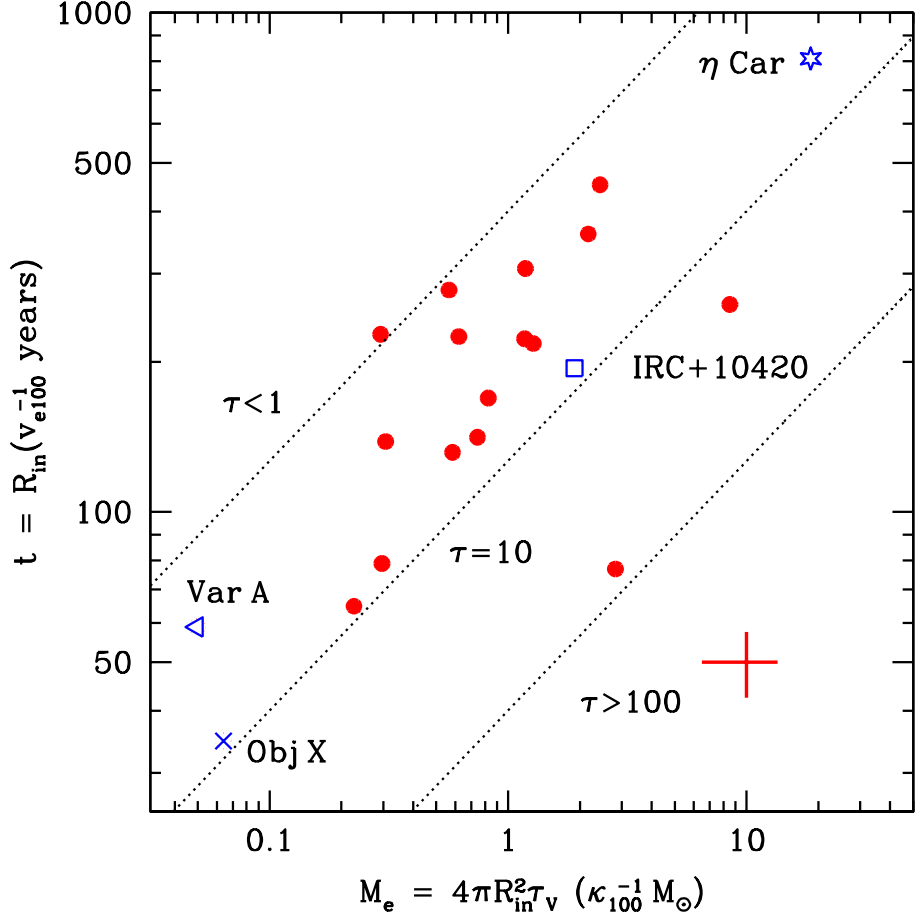}
\end{center}
\caption{Elapsed time $t= R_{in} {v_{e100}^{\,\,-1}}$ 
as a function of the estimated ejecta mass $M_e$
for the best fit graphitic models.
The mass and radius are scaled to 
$\kappa_V=100\,\kappa_{100}$\,cm$^2$\,gm$^{-1}$ and $v_e=100\,v_{e100}$\,km\,s$^{-1}$, and 
can be rescaled as $t\propto{v_e^{-1}}$ and 
$M_e\propto{\kappa_V^{-1}}$.
The error bar shows the typical $1\sigma$ uncertainties on $t$ ($\pm 15\%$)
and $M_e$ ($\pm 35\%$) of the best SED fit models..
The three dotted lines correspond to optical depths
$\tau_V=1,10$ and $100$.
We should have trouble finding sources with $\tau_V<1$ due to 
lack of mid-IR emission and $\tau_V\gtrsim100$ due to
the dust photosphere being too cold (peak emission in far-IR).
The large $t$ estimate for $\eta$\,Car when 
scaled by $v_{e100}$ is due to the anomalously 
large ejecta velocities ($\sim600$km\,s$^{-1}$ along the 
long axis \citep{ref:Cox_1995,ref:Smith_2006a} compared to 
typical LBV shells ($\sim50$km\,s$^{-1}$, \citealp{ref:Tiffany_2010}).}
\label{fig:tv}
\end{figure}

One check on our selection methods is to examine the distribution of shell radii.
Crudely, we can see a shell until 
it either becomes optically thin or too cold, so the probability 
distribution of a shell's radius assuming a constant expansion velocity is 
\begin{equation}
\frac{dN}{dR_{in}} = \frac{1}{R_{max}} = \hbox{constant}
\end{equation}
for $R_{in}<R_{max}$. An ensemble of shells with similar $R_{max}$ should 
then show this distribution.
Figure\,\ref{fig:complete} shows the cumulative histogram (excluding N\,7793-3) 
of the inner shell radii ($R_{in}$).
The curves show the expected distribution where we simply normalized to the 
point where $F\left(<R_{in}\right)\simeq0.5$. The agreement shows that our sample 
should be relatively complete up to $R_{max}\simeq10^{16.5}$-$10^{17}$\,cm
which corresponds to a maximum age of 
\begin{equation}
\label{eqn:rad}
t_{max} \simeq 300\,{v_{e100}^{\,\,-1}}\,\,\,\hbox{years}. 
\end{equation}
Figure\,\ref{fig:tv} shows the age ($t_e=R_{in}/{{v_{e100}^{-1}}}$) of the shells 
as a function of $M_e$.
We also show lines corresponding to optical depths of
$\tau_V=1,10,100$. As expected, we see no sources with very low or high 
optical depths, as we should have trouble finding sources with $\tau_V<1$ due to a
lack of mid-IR emission and $\tau_V\gtrsim100$ due to
the dust photosphere being too cold (peak emission in the far-IR).
Indeed, most of the dusty stars have $ 1 < \tau_V < 10 $ and 
none has $ \tau_V > 100$. The large $t_e$ estimate for $\eta$\,Car when 
scaled by $v_{e100}$ is due to its unusually 
large ejecta velocities ($\sim600$km\,s$^{-1}$ along the 
long axis \citep{ref:Cox_1995,ref:Smith_2006a} compared to 
typical LBV shells ($\sim50-100$km\,s$^{-1}$, \citealp{ref:Tiffany_2010}).

\section{Implications}
\label{sec:discussion3}

The advantage of surveying external galaxies with a significant supernova
rate is that we can translate our results into estimates of abundances and
rates. We scale our rates using the observed supernova rate of 
$R_{SN}=0.15$~year$^{-1}$ ($0.05 < R_{SN} < 0.35$ at 90\% confidence).
As we discussed in Paper\,I, this is significantly higher than standard star
formation rate estimates for these galaxies, but the SN rate is directly
proportional to the massive star formation rate rather than an indirect
indicator, and similar discrepancies, although not as dramatic, have been 
noted in other contexts \citep[e.g.,][]{ref:Horiuchi_2011}.
In this section we first outline how we will
estimate rates, and then we discuss the constraints on analogs of $\eta$\,Car
and the implications of our sample of luminous dusty stars.

We are comparing a sample of $N_{SN}=3$ supernovae observed over $t_{SN}=20$~years
to a sample of $N_c$ candidate stars which are detectable by our selection procedures
for a time $t_d$. In Paper\,I we used DUSTY to model the detection of expanding 
dusty shells and found that a good estimate for the detection time period was
\begin{equation}
 t_d = t_w + 66 \left( { 100~\hbox{km~s}^{-1} \over v_e } \right)
 \left( { L_* \over 10^6 L_\odot } \right)^{0.82}
 \left( { M_e \over M_\odot } \right)^{0.043}~\hbox{years}
\end{equation}
for shells with masses in the range $-1 \leq \log M_e/M_\odot \leq 1$ around
stars of luminosity $5.5 \leq \log L_*/L_\odot \leq 6.5$ where $t_w$ is the
duration of the ``wind'' phase and the second term is an estimate of how long
the shell will be detected after the heavy mass loss phase ends. The principle
uncertainty lies in the choice of the velocity, $v_e$. If the rate of events
in the sample is $R_e$, then we expect to find $N_e = R_e t_d$ candidates. 

The transient rate in a sample of galaxies is less interesting than comparing the rate
to the supernova rate. Let $f_e$ be the fraction of massive ($M_{ZAMS}>8\,M_\odot$) stars that
create the transients, where $f_e = (M_C/8M_\odot)^{-1.35}$ if we 
assume a Salpeter IMF \citep{ref:Kennicutt_1998b}, that all stars more massive than $8M_\odot$ become
supernovae and that all stars more massive than $M_C$ cause the transients.
If each star undergoes an average of $N_e$ eruptions, then the rate of transients is related
to the rate of supernovae by $ R_e = N_e f_e R_{SN} = F_e R_{SN}$. The
interesting quantity to constrain is $F_e = N_e f_e$ rather than $R_e$. Poisson
statistics provide constraints on the rates, 
where $P(D|R) \propto (R t)^N \exp(-R t)$ for $N$ events observed over a time
period $t$. This means that the probability of the rates given the data is
\begin{equation}
 P(R_{SN},R_e|D) \propto P(R_{SN}) P(R_e) (R_{SN} t_{SN})^3 (R_e t_d)^{N_c} 
 \exp(-R_{SN}t_{SN}-R_e t_d)
\end{equation}
where $P(R_{SN})$ and $P(R_e)$ are priors on the rates which we will assume to
be uniform and we have set $N_{SN}=3$. If we now change variables to compute
$F_e$ and marginalize over the unknown supernova rate, we find that the 
probability distribution for the ratio of the rates is
\begin{equation}
 P(F_e |D) \propto F_e^{N_c} \left( F_e t_d + t_{SN}\right)^{-5-N_c}
\end{equation}
with the standard normalization that $\int P(F_e|D) dF_e \equiv 1$.
For our estimates of $F_e$ we present either 90\% confidence upper limits
or the value corresponding to the median probability and symmetric 90\% probability
confidence regions. Note that the probability distribution really just depends
on the product $F_e t_d$, so the results for any given estimate of $t_d$ 
are easily rescaled.

\subsection{No $\eta$\,Car Analog Is Found}

It is immediately obvious from Figure\,\ref{fig:HR_like} that none of the sources we identified closely
resemble $\eta$\,Car. Their typical luminosities of $10^{5.7\pm0.2}L_\odot$ correspond
to $\sim 40M_\odot$ stars 
\citep{ref:Maeder_1981,ref:Maeder_1987,ref:Maeder_1988,ref:Stothers_1996,ref:Meynet_1994} 
rather than the higher masses usually associated with
LBV outbursts. Since we identify a significant population of fainter stars, this is 
unlikely to be a selection effect, and we conclude that these galaxies contain no analogs 
of $\eta$\,Car.

There are two ways we can interpret the result. First, we
can ignore the existence of $\eta$\,Car, and set $N_c=0$. Alternatively, we can
acknowledge the existence of $\eta$\,Car, in which case $N_c=1$, since $\eta$\,Car 
passes our selection criterion and mid-IR surveys of our Galaxy for objects as luminous
as $\eta$\,Car are probably complete.
For the first case, the 90\% confidence upper limit is 
$F_e < 0.077 t_{d200}^{-1}$ where the period over which such
systems can be detected is scaled to $t_d = 200 t_{d200}$~years.
For the second case, where we include $\eta$\,Car, we find that
$F_e = 0.046 t_{d200}^{-1}$ with $0.0083 < F_e t_{d200} < 0.19$ at 90\% confidence.
In either case, the rate of transients comparable to $\eta$\,Car is a small
fraction of the supernova rate.

Stars as massive as $\eta$\,Car are also rare, representing only $f_e=0.02$
to $0.04$ of all massive stars for a mass range from $70$/$100M_\odot$ to
$200M_\odot$. If every sufficiently massive star had one eruption, the
results including $\eta$\,Car correspond to a minimum mass of $M_C = 65M_\odot$ 
($26 M_\odot < M_C < 138 M_\odot$). If every star has an
average of two eruptions, the mass limits rise to $M_C = 94M_\odot$ 
($42 M_\odot < M_C < 162 M_\odot$). Similarly the upper limit from
ignoring the existence of $\eta$\,Car corresponds to $M_C > 48M_\odot$
for an average of one eruption or $M_C > 72 M_\odot$ for an average
of two. \cite{ref:Kochanek_2011c} estimated that the abundance of lower optical
depth shells found at $24\mu$m around massive stars in the Galaxy was
roughly consistent with all stars more massive than $M_C = 40M_\odot$
having an average of two eruptions, corresponding to $F_e \simeq 0.2$,
which is consistent with the present results, but close to the upper limits.

\subsection{An Emerging Class of Dust Obscured Stars}

All the newly identified stars have luminosities within a narrow range of
$\log L/L_\odot \simeq 5.5$-$6.0$ (see Figure\,\ref{fig:Dusty}), which roughly corresponds
to initial stellar masses of $M_{ZAMS}\simeq 25$-$60 M_\odot$ 
\citep[see Section 4 of][and references therein]{ref:deJager_1998}.
Local examples of evolved stars in this luminosity range are the 
Yellow Hypergiants (YHGs) such as IRC$+$10420, $\rho$\,Cas and 
HR\,8752 \citep{ref:deJager_1997,ref:Smith_2006},
many of which are also partially obscured by dust ejecta. There is
no means of cleanly surveying the Galaxy for these objects and they 
are so rare that samples in the Galaxy and the Magellanic Clouds 
do not provide good statistics for their abundances, life times or
total mass loss. Our well-defined sample of likely extragalactic 
analogs provides a means of addressing some of these questions.

If we assume these objects are similar to stars like IRC$+$10420, their
expansion velocities will be more like $50$~km/s than the $100$~km/s
of the typical LBV shell. Hence, it seems more appropriate to scale
the results to $t_d = 500 t_{d500}$~years. This also matches the
estimated age of the phase of dusty mass loss by IRC$+$10420 \citep{ref:Tiffany_2010}.
With 18 candidates, this detection period 
then leads to a median estimate that $F_e = 0.20 t_{d500}^{-1}$ with
$0.086 < F_e t_{d500} < 0.55$. If we associate these with the mass
range from $25$ to $60M_\odot$, they represent a fraction of 
$f_e \simeq 0.15$ of massive stars, so the average number of
episodes per star, $N_e=F_e/f_e \simeq 1.3 t_{d500}^{-1}$ with
a possible range of $0.58 < N_e t_{d500} < 3.7$, although this
does not include the uncertainties in $f_e$

Figure\,\ref{fig:HR_like} shows that the median mass causing the obscuration is
$M_e \sim 0.5 M_\odot$. The total mass lost in all the eruptions is then of order
$N_e M_e $, which would be of order $0.3$-$1.9 t_{d500}^{-1} M_\odot$. 
This implies that the periods of optically thick (dusty) mass loss cannot 
dominate the overall
mass loss of the star. To make the mass lost in these phases dominate
either requires that we have grossly overestimated $t_d$, or that
the mass range of the stars is much narrower.
A related point is that these phases represent a negligible fraction
of the post-main-sequence life times of the stars, at most lasting
a few thousand years.

\section{Conclusions}
\label{sec:conclusions}

In our survey, we have found no true analogs of $\eta$\,Car.
This implies that the rate of Great Eruption-like
events is of order $F_e = 0.046 t_{d200}^{-1}$ ($0.0083 < F_e t_{d200} < 0.19$)
of the ccSN rate, which is roughly consistent with each $M \gtrsim 70M_\odot$
star undergoing 1 or 2 such outbursts in its lifetime. This is scaled by
an estimated detection period of order $t_d =200 t_{d200}$\,years. 
We do identify a significant population of lower luminosity dusty stars that
that are likely similar to IRC$+10420$.
Stars enter this phase at the rate $F_e = 0.20 t_{d500}^{-1}$
($0.086 < F_e t_{d500} < 0.55$) compared to the ccSN rate and for a 
detection period of $t_d = 500 t_{d500}$\,years. Here the
detection period is assumed longer because the expansion velocities
are likely slower. This rate is comparable to 
having all stars with $25 < M < 60M_\odot$ undergoing such a phase 
once or twice. 

\begin{figure}
\begin{center}
\includegraphics[angle=0,width=130mm]{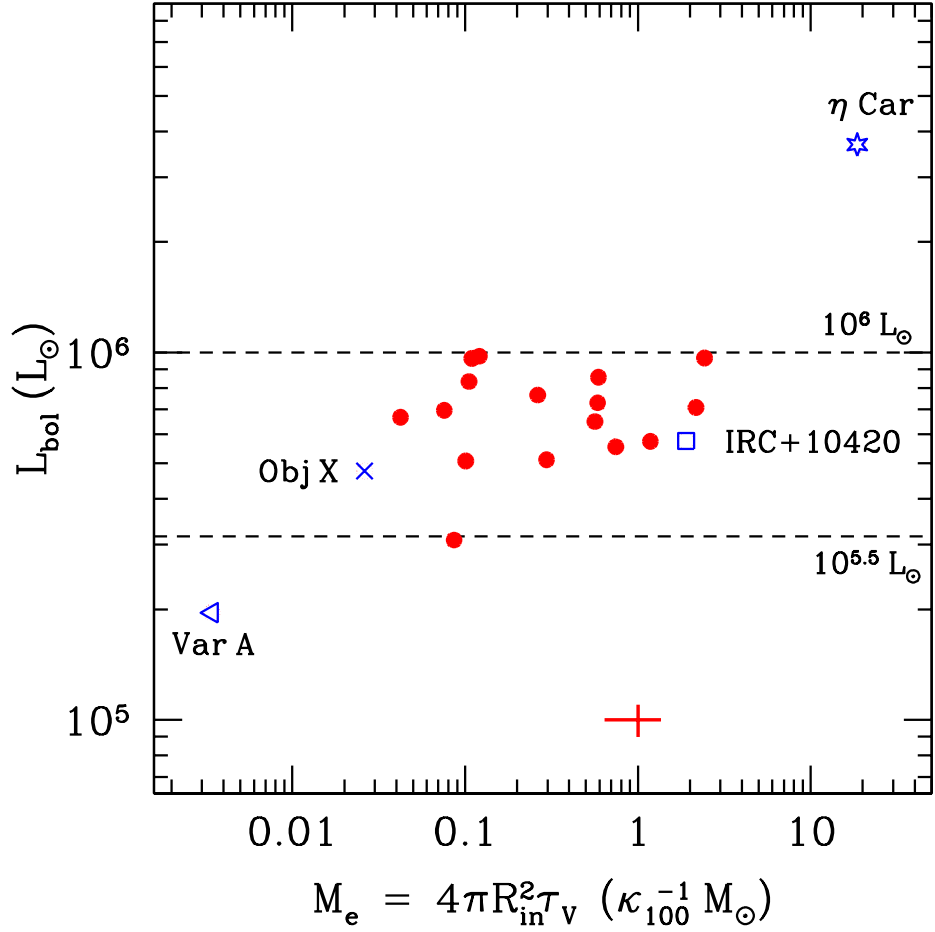}
\end{center}
\caption{Luminosities of the obscured stars
as a function of the estimated ejecta mass determined
from the best fit model for each SED.
The dashed lines enclose the luminosity range $\log(L/L_{sun})\simeq5.5-6.0$.
We do not show N\,7793-3 for which we have no optical or near-IR data.
$IRC+10420$ (square), M\,33\,Var\,A (triangle), and $\eta$\,Car (star symbol) 
are shown for comparison. The error bar 
corresponds to the typical $1\sigma$ uncertainties on $L_{bol}$ ($\pm 10\%$) 
and $M_e$ ($\pm 35 \%$) of the best SED fit models.}
\label{fig:HR_like}
\end{figure}

\begin{figure}[p]
\begin{center}
\includegraphics[angle=0,width=160mm]{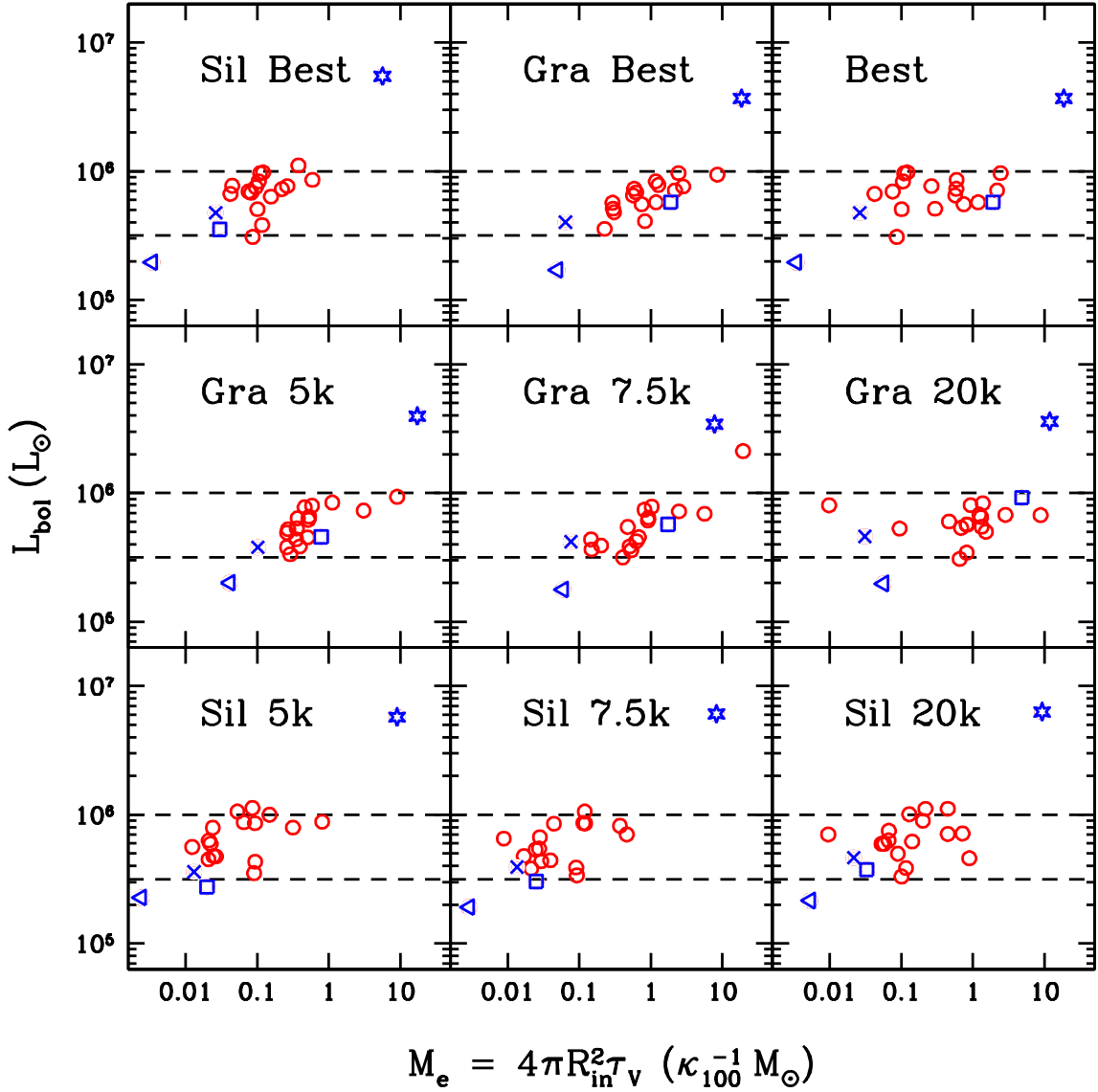}
\end{center}
\caption{Same as Figure\,\ref{fig:HR_like}, but for different dust types and 
temperature assumptions. The top row shows the best silicate (left), graphitic (center), 
and the better of the two (right, same as Figure\,\ref{fig:HR_like}) models.
The middle and bottom rows show the best fit models for graphitic 
and silicate dust at fixed stellar temperatures of 5000\,K, 7500\,K and 20000\,K.
The only higher luminosity case in the fixed temperature model panels is N\,2403-4,
for which the best fit models have significantly smaller $\chi^2$ and lower luminosities
for both dust types.}
\label{fig:Dusty}
\end{figure}

If the estimated detection periods and mass ranges are roughly correct, 
and our completeness is relatively high, there are two interesting 
implications for both populations. First, these high optical depth
phases represent a negligible fraction of the post-main sequence 
lifetimes of these stars, at most lasting a few thousand years. 
This implies that these events have to be associated with special
periods in the evolution of the stars. The number of such events a 
star experiences is also small, one or two, not ten or twenty. 
Second, while a significant amount of mass is lost in the eruptions,
they cannot be a dominant contribution to mass loss. For these
high mass stars, standard models 
\citep[e.g.,][]{ref:Maeder_1981,ref:Maeder_1987,ref:Maeder_1988,ref:Stothers_1996,ref:Meynet_1994} 
typically strip the stars of their hydrogen envelopes and beyond,
implying total mass losses of all but the last $5$-$10M_\odot$.
The median mass loss in Figure\,\ref{fig:HR_like} is $M_e \sim 0.5 M_\odot$ and
if every star underwent two eruptions, the typical total would 
be $N_e M_e \sim M_\odot$. Clearly there are some examples that
require significantly larger $M_e$, but we simply do not find 
enough heavily obscured stars for this phase to represent more
than a modest fraction of the total mass loss ($\sim 10\%$ not $\sim 50\%$). 

For the stars similar to IRC$+$10420, this is consistent with the
picture that the photospheres of blue-ward evolving Red Super Giants (RSGs) with
$\log(L_*/L_\odot)= 5.6\sim6.0$ and $T_{star} \simeq 7000$-$12500$~K,
become moderately unstable, leading to periods of 
lower effective temperature and enhanced mass loss as the stars try 
to evolve into a ``prohibited'' region of the HR diagram that 
\citet{ref:deJager_1997}, 
\citet{ref:deJager_1998} and \citet{ref:Nieuwenhuijzen_2000}
termed the ``yellow void''.
In this phase, the stars lose enough mass to evolve into a hotter, 
less massive star on the blue side of the HR diagram. This is also
the luminosity regime of the ``bistability jump'' in wind speeds
driven by opacity changes which \citet{ref:Smith_2004} hypothesizes
can explain the absence of LBVs and the existence of YHGs with high
mass loss rates and dust formation (\citealt{ref:Vink_2009}) in this luminosity range. 
In fact, \citet{ref:Humphreys_2002} propose that IRC$+10420$,
which is identified by our selection criterion, is such a star.
While these arguments supply a unique, short-lived evolutionary 
phase, there may be problems with the absolute scale of the mass
loss, since estimates are that IRC$+$10420 started with a mass
of $\sim 40 M_\odot$ and has lost all but $6\sim15 M_\odot$
\citep{ref:Nieuwenhuijzen_2000}. 

The only other similarly unique phase in the lives of these stars is the final
post-carbon ignition phase. There are now many examples of stars
which have had outbursts shortly before exploding as supernovae
\citep[e.g.,][]{ref:Pastorello_2007,ref:Mauerhan_2012,ref:Prieto_2012,ref:Pastorello_2012,ref:Ofek_2013a} 
and superluminous supernovae that are most
easily explained by surrounding the star with a large amount of
previously ejected mass 
\citep{ref:Smith_2007b,ref:GalYam_2007,ref:Smith_2008,ref:Kozlowski_2010,ref:Ofek_2013a}. 
Powering these supernovae requires mass ejected in 
the last years to decades of the stellar life 
\citep[e.g.,][]{ref:Chevalier_1994,ref:Chugai_2003,ref:Smith_2009,ref:Moriya_2014}. 
and it seems natural to associate these events with the mass ejections of LBVs like
$\eta$\,Car \citep[e.g.,][]{ref:Smith_2007b,ref:GalYam_2009}. The 
statistical properties and masses of either of the classes of dusty
stars we discuss are well-matched to the statistical requirements for
explaining these interaction powered supernovae if the instability is 
associated with the onset of carbon burning \citep[see][]{ref:Kochanek_2011c}. 
If there is only one eruption mechanism, it must be associated with a relatively
long period like carbon burning (thousands of years) rather than
the shorter, later nuclear burning phases, because we observe many
systems like $\eta$\,Car that have survived far longer than these
final phases last. If the mechanism for producing the ejecta around
the superluminous supernovae is associated with nuclear burning phases
beyond carbon, then we must have second eruption mechanism to explain
$\eta$\,Car or other still older LBVs surrounded by massive dusty
shells. If there indeed are two mass loss mechanisms --- one commencing 
$\gtrsim 10^3$\,years from core-collapse and the other occurring in 
the $\sim 1$\,year prior to
core-collapse --- then the self-obscured stars identified in this 
work may very well be experiencing the earlier of these two mechanisms.
Otherwise, in a larger sample to $\sim100$ such stars, one should be 
exploding as a ccSN every $\sim10$ years.

The dusty stars can be further characterized by their variability,
which will help to follow the evolution of the dust. For the 
optically brighter examples, it may be possible to spectroscopically
determine the stellar temperatures, although detailed study may
only become possible with the James Webb Space Telescope (\textit{JWST}).
It is relatively easy to expand our survey to additional galaxies. 
For very luminous sources like $\eta$\,Car analogs this is probably
feasible to distance of $10$\,Mpc, while for the lower luminosity 
IRC$+$10420 analogs this is likely only feasible at the distances
of the most distant galaxies in our sample ($\sim4$\,Mpc). Larger galaxy
samples are needed not only to increase the sample of dusty luminous 
stars (and hopefully find a true $\eta$ Car analog!), but also to
have a sample with a larger number of supernovae, or equivalently
a higher star formation rate. Our estimate of the abundance of 
IRC$+$10420 analogs is limited by the small number of ccSN
($3$) in our sample more than by the number of dusty stars ($18$) identified. 
Finally, while we have shown that surveys for the stars are feasible
using archival Spitzer data, JWST will be a far more powerful probe
of these stars. The HST-like resolution \citep{ref:Gardner_2006} 
will be enormously useful
to either greatly reduce the problem of confusion or to greatly expand
the survey volume. Far more important will be the ability to carry
out the survey at $24\mu$m, which will increase the time over which
dusty shells can be identified from hundreds of years to thousands
of years, greatly improving the statistics and our ability to survey
the long term evolution of these systems and the relationship between
stellar eruptions and supernovae.

\acknowledgments
We thank Hendrik Linz for helping us analyze the Herschel PACS data and John Beacom for 
numerous productive discussions.
We extend our gratitude to the SINGS Legacy Survey and LVL Survey for making
their data publicly available. This research has made use of 
observations made with the NASA/ESA Hubble Space Telescope, 
and obtained from the Hubble Legacy Archive, which is a collaboration 
between the Space Telescope Science Institute (STScI/NASA), the Space 
Telescope European Coordinating Facility (ST-ECF/ESA) and the 
Canadian Astronomy Data Centre (CADC/NRC/CSA).
RK and KZS are supported in part by NSF grant AST-1108687.

\clearpage

\clearpage

\clearpage

\clearpage

\clearpage

\clearpage

\clearpage

\clearpage

\clearpage

\clearpage

\clearpage

\clearpage

\clearpage

\begin{table}   
\begin{center}   
\caption{PACS Aperture Definitions}
\label{tab:pacs}
\begin{tabular}{rrrrrr}
\\   
\hline  
\hline 
\multicolumn{1}{c}{Band ($\micron$)} &
\multicolumn{1}{c}{Pixel Scale} &
\multicolumn{1}{c}{$R_{ap}$} &
\multicolumn{1}{c}{$R_{in}$} &
\multicolumn{1}{c}{$R_{out}$} &
\multicolumn{1}{c}{Ap. Corr.}
\\
\hline  
\hline
\\
$70\,\micron$ & $3\farcs2$ & $6\farcs4$ & $60\farcs8$ &  $70\farcs4$  & ${0.72}^{-1}$ \\
$100\,\micron$   & $3\farcs2$ & $6\farcs4$ & $60\farcs8$ &  $70\farcs4$  & ${0.69}^{-1}$ \\
$160\,\micron$   & $6\farcs4$ & $12\farcs8$   & $121\farcs6$   &  $140\farcs8$ & ${0.78}^{-1}$ \\
\\
\hline  
\hline  
\end{tabular} 
\end{center}  
\end{table}

\clearpage

\begin{landscape}
\begin{table*}[p]   
\begin{center}   
\begin{tiny}  
\caption{Multi-Wavelength Photometry\tablenotemark{a}} 
\label{tab:photo}   
\begin{tabular}{rrrrrrrrrrrrrrrrrr}   
\\   
\hline  
\hline  
\multicolumn{1}{c}{ID} &  
\multicolumn{1}{c}{$U$} & 
\multicolumn{1}{c}{$B$} & 
\multicolumn{1}{c}{$V$} & 
\multicolumn{1}{c}{$R$} & 
\multicolumn{1}{c}{$I$} & 
\multicolumn{1}{c}{$J$} & 
\multicolumn{1}{c}{$H$} & 
\multicolumn{1}{c}{$K_s$} &  
\multicolumn{1}{c}{[3.6]} &  
\multicolumn{1}{c}{[4.5]} &  
\multicolumn{1}{c}{[5.8]} &  
\multicolumn{1}{c}{[8.0]} &  
\multicolumn{1}{c}{[12]} &   
\multicolumn{1}{c}{[24]} &   
\multicolumn{1}{c}{[70]} &   
\multicolumn{1}{c}{[100]} &  
\multicolumn{1}{c}{[160]} 
\\   
\hline  
\hline
\\
M\,$33-1$ & $<24.1$ & $<24.3$ & $23.15$ & $21.61$ & $19.99$ & $17.07$ & $15.04$ & $13.60$ & $11.73$ & $10.92$ & $9.97$ & $9.08$ & $7.85$ & $5.71$ & $278.1$ & $377.1$ & $1181$ \\  
M\,$33-2$ & $20.61$ & $21.84$ & $21.14$ & $20.13$ & $20.48$ & $15.96$ & $14.68$ & $14.02$ & $11.93$ & $11.56$ & $9.91$ & $8.75$ & \dots  & $3.27$ & $4503$ & $5791$ & $7501$ \\ 
M\,$33-3$ & $18.00$ & $19.26$ & $19.10$ & $18.16$ & $18.92$ & \dots & \dots & \dots  & $12.39$ & $12.11$ & $9.80$ & $8.56$ & $7.35$ & $4.75$ & $1064$ & $1407$ & $1916$ \\   
M\,$33-4$ & $19.14$ & $20.37$ & $20.00$ & $19.08$ & $19.62$ & \dots & \dots & \dots  & $12.66$ & $12.54$ & $10.07$ & $8.71$ & $6.84$ & $4.44$ & $1648$ & $2249$ & $4017$ \\  
M\,$33-5$ & $16.84$ & $17.43$ & $16.62$ & $16.62$ & $17.12$ & $15.77$ & $15.17$ & $14.28$ & $11.61$ & $10.51$ & $8.91$ & $7.20$ & $5.06$ & $1.38$ & $8959$ & $7250$ & $5237$ \\ 
M\,$33-6$ & $22.17$ & $22.73$ & $21.23$ & $20.04$ & $20.42$ & \dots & \dots & \dots  & $12.33$ & $11.83$ & $10.37$ & $8.43$ & \dots  & $3.03$ & $3026$ & $3338$ & $3250$ \\  
M\,$33-7$ & $20.09$ & $20.84$ & $20.25$ & $19.45$ & $19.48$ & \dots & \dots & \dots  & $12.68$ & $12.26$ & $9.89$ & $8.52$ & $6.91$ & $4.22$ & $1548$ & $4955$ & $2329$ \\   
M\,$33-8$ & $18.81$ & $19.85$ & $18.81$ & $17.94$ & $17.68$ & $16.15$ & $15.49$ & $14.25$ & $11.52$ & $11.01$ & $8.78$ & $7.25$ & $5.44$ & $2.38$ & $4955$ & $5581$ & $5607$ \\ 
M\,$33-9$ & $19.60$ & $20.42$ & $19.81$ & $19.22$ & $19.80$ & \dots & \dots & \dots  & $12.74$ & $12.31$ & $10.33$ & $8.48$ & \dots  & $3.24$ & $2875$ & $3590$ & $3206$ \\  
N\,$300-1$ & \dots  & $25.23$ & $23.68$ & \dots  & $21.13$ & \dots & \dots & \dots  & $13.22$ & $12.23$ & $11.22$ & $9.91$ & $8.33$ & $6.90$ & $0.96$\tablenotemark{r} & $200.1$ & $475$ \\ 
N\,$2403-1$ & \dots  & $9.23$ & \dots  & $11.04$ & $12.28$ & $14.89$ & $14.42$ & $14.21$ & $14.10$ & $13.79$ & $12.60$ & $10.54$ & $7.93$ & $6.15$ & $465.4$ & \dots  & $1385$ \\  
N\,$2403-2$ & \dots  & $21.3$ & $21.3$ & \dots  & $21.3$ & \dots & \dots & \dots  & $14.77$ & $14.67$ & $12.47$ & $10.91$ & \dots  & $7.77$ & $330.4$ & \dots  & $2045$ \\   
N\,$2403-3$ & $<19.5$ & $<20.0$ & $<19.9$ & $<19.6$ & \dots & \dots & \dots & \dots  & $15.22$ & $14.28$ & $12.51$ & $10.65$ & $8.98$ & $7.22$ & $148.4$ & \dots  & $790.5$ \\  
N\,$2403-4$ & \dots  & $25.6$ & $23.5$ & \dots  & $20.5$ & $17.21$ & $16.05$ & $14.45$ & $14.65$ & $14.60$ & $12.50$ & $10.59$ & \dots  & $7.94$ & $408.8$ & \dots  & $2100$ \\ 
N\,$2403-5$ & \dots  & $21.1$ & $20.3$ & \dots  & $19.7$ & \dots & \dots & \dots  & $14.64$ & $14.13$ & $12.74$ & $10.41$ & \dots  & $7.45$ & $369.4$ & \dots  & $2230$ \\   
M\,$81-5$ & \dots  & $<25$ & $<25$ & \dots & \dots & \dots & \dots & \dots  & $14.93$ & $14.25$ & $13.16$ & $11.99$ & $9.86$ & $8.36$ & $32.7$ & \dots  & $346.7$ \\   
M\,$81-6$ & \dots  & $<25$ & $<24.5$ & \dots & \dots & \dots & \dots & \dots  & $15.26$ & $14.18$ & $13.09$ & $12.03$ & $10.26$ & $8.72$ & $30.1$ & \dots  & $142.1$ \\   
M\,$81-7$\tablenotemark{b} & $17.55$ & $17.57$ & $17.09$ & $17.46$ & $17.66$ & \dots & \dots & \dots  & $14.89$ & $14.07$ & $13.19$ & $11.78$ & $10.15$ & $8.15$ & $72.4$ & \dots  & $343.8$ \\   
M\,$81-10$ & \dots  & $19.25$ & $19.20$ & \dots & \dots & \dots & \dots & \dots  & $14.00$ & $13.15$ & $12.13$ & $10.02$ & $8.00$ & $5.80$ & $589.2$ & \dots  & $1534$ \\ 
M\,$81-11$ & \dots  & $22.10$ & $21.10$ & \dots  & $19.83$ & \dots & \dots & \dots  & $15.09$ & $14.50$ & $13.17$ & $11.22$ & \dots  & $8.42$ & $142.3$ & \dots  & $1067$ \\ 
M\,$81-12$ & \dots  & $23.95$ & $21.98$ & \dots  & $19.07$ & \dots & \dots & \dots  & $15.70$ & $15.10$ & $13.10$ & $11.31$ & \dots  & $7.79$ & $275.6$ & \dots  & $1141$ \\ 
M\,$81-14$ & \dots  & $<24.5$ & $<24$ & \dots  & $<23.5$ & \dots & \dots & \dots  & $15.61$ & $15.30$ & $13.01$ & $8.74$ & \dots  & $7.31$ & $243.1$ & \dots  & $966.3$ \\   
N\,$247-1$ & \dots & \dots & \dots & \dots & \dots & \dots & \dots & \dots  & $15.04$ & $13.86$ & $12.87$ & $11.56$ & $10.51$ & $8.23$ & \dots & \dots  & $1.84$\tablenotemark{r} \\  
N\,$247-3$ & \dots  & $15.73$ & \dots  & $15.87$ & \dots & $15.73$ & $14.79$ & $14.58$ & $14.80$ & $14.20$ & $14.38$ & $10.80$ & $9.88$ & $8.14$ & $2.98$\tablenotemark{r} & \dots  & $-0.59$\tablenotemark{r} \\   
N\,$7793-1$ & \dots & \dots & \dots & \dots & \dots & \dots & \dots & \dots  & $14.72$ & $13.79$ & $13.85$ & $11.74$ & $10.79$ & $8.65$ & $4.49$\tablenotemark{r} & $5.142$ & $-0.69$\tablenotemark{r} \\  
N\,$7793-3$ & \dots & \dots & \dots & \dots & \dots & \dots & \dots & \dots  & $14.89$ & $14.64$ & $13.42$ & $11.89$ & $9.92$ & $8.67$ & $57.9$ & $89.61$ & $228.2$ \\ 
N\,$7793-4$ & \dots & \dots & \dots & \dots & \dots  & $16.30$ & $15.59$ & $15.17$ & $14.40$ & $13.70$ & $12.97$ & $11.08$ & $9.77$ & $7.85$ & $63.7$ & $58.18$ & $76.83$ \\ 
N\,$7793-6$ & \dots & \dots  & $19.5$ & \dots  & $18.5$ & $16.45$ & $15.98$ & $15.58$ & $15.09$ & $14.88$ & $13.19$ & $11.47$ & $9.09$ & $8.12$ & $152.1$ & $267.7$ & $1044$ \\ 
N\,$7793-8$ & \dots & \dots & \dots & \dots & \dots  & $16.07$ & $15.56$ & $14.14$ & $14.72$ & $14.93$ & $13.27$ & $11.08$ & $9.67$ & $8.58$ & $50.9$ & $111.8$ & $207.2$ \\ 
N\,$7793-9$ & \dots  & $25.7$ & $23.0$ & \dots  & $22.3$ & \dots & \dots & \dots  & $15.65$ & $15.38$ & $13.40$ & $11.57$ & \dots  & $8.64$ & $103.6$ & $210.1$ & $989.7$ \\ 
N\,$7793-10$ & \dots  & $23.0$ & $21.2$ & \dots  & $19.6$ & \dots & \dots & \dots  & $15.26$ & $15.29$ & $13.00$ & $11.23$ & $9.62$ & $7.83$ & $114.0$ & $205.1$ & $426.1$ \\   
N\,$7793-11$ & \dots  & $19.6$ & $19.5$ & \dots  & $19.5$ & \dots & \dots & \dots  & $15.51$ & $15.20$ & $12.98$ & $11.20$ & $8.98$ & $7.20$ & $213.5$ & $349.9$ & $669.1$ \\   
N\,$7793-12$ & \dots & \dots & \dots & \dots & \dots & \dots & \dots & \dots  & $15.89$ & $15.37$ & $13.58$ & $11.57$ & \dots  & $7.84$ & $63.3$ & $121.0$ & $666.7$ \\   
N\,$7793-13$ & \dots  & $22.4$ & $21.9$ & \dots  & $22.1$ & \dots & \dots & \dots  & $15.92$ & $15.39$ & $12.72$ & $11.15$ & \dots  & $7.25$ & $176$ & $345.4$ & $1141$ \\   
N\,$7793-14$ & \dots  & $20.0$ & $19.0$ & \dots  & $16.6$ & \dots & \dots & \dots  & $15.77$ & $15.47$ & $13.02$ & $11.28$ & $9.58$ & $7.97$ & $111.4$ & $185.2$ & $614.8$ \\ 
\\
\hline  
\hline  
\end{tabular} 
\end{tiny} 
\end{center}  
\tablenotetext{a}{Optical, near-IR, \textit{Spitzer} IRAC $3.6-8.0\,\micron$, WISE $12\,\micron$, and \textit{Spitzer} MIPS 
$24$, $70$, and $160\,\micron$ measurements in apparent magnitudes. \textit{Herschel} PACS  $70$, $100$ and $160\,\micron$ 
measurements in flux (mJy). WISE, MIPS, and PACS measurements are always treated as upper limits.}
\tablenotetext{b}{Optical measurement are SDSS $ugriz$ magnitudes, not $UBVRI$.}   
\tablenotetext{r}{\textit{Spitzer} MIPS $70\,\micron$ and $160\,\micron$ apparent magnitudes, not \textit{Herschel} flux (mJy).} 
\end{table*}
\end{landscape}

\clearpage

\begin{landscape}
\begin{table*}[p]   
\begin{center}   
\begin{tiny}  
\caption{Best Fit Models}  
\label{tab:mcmc}   
\begin{tabular}{rrrrrrrrrrrrrrrrrrrrr}   
\\   
\hline  
\hline  
\multicolumn{3}{c}{} &  
\multicolumn{8}{c}{\textit{Graphitic}} &  
\multicolumn{2}{c}{} & 
\multicolumn{8}{c}{\textit{Silicate}} 
\\   
\multicolumn{1}{c}{ID} &  
\multicolumn{1}{c}{} &
\multicolumn{1}{c}{} &
\multicolumn{1}{c}{$\chi^2$} & 
\multicolumn{1}{c}{$\tau_V$} & 
\multicolumn{1}{c}{$T_d$} & 
\multicolumn{1}{c}{$T_*$} & 
\multicolumn{1}{c}{$\log\left(R_{in}\right)$} & 
\multicolumn{1}{c}{$\log L_*$} &
\multicolumn{1}{c}{$M_e$} &  
\multicolumn{1}{c}{$t_e$} &
\multicolumn{1}{c}{} &
\multicolumn{1}{c}{} &
\multicolumn{1}{c}{$\chi^2$} & 
\multicolumn{1}{c}{$\tau_V$} & 
\multicolumn{1}{c}{$T_d$} & 
\multicolumn{1}{c}{$T_*$} & 
\multicolumn{1}{c}{$\log\left(R_{in}\right)$} &  
\multicolumn{1}{c}{$\log L_*$} &
\multicolumn{1}{c}{$M_e$} &
\multicolumn{1}{c}{$t_e$} 
\\   
\multicolumn{5}{c}{} &  
\multicolumn{1}{c}{(K)} & 
\multicolumn{1}{c}{(K)} & 
\multicolumn{1}{c}{(cm)} &
\multicolumn{1}{c}{($L_\odot$)} &
\multicolumn{1}{c}{($M_\odot$)}  & 
\multicolumn{1}{c}{(years)}  &
\multicolumn{4}{c}{} &
\multicolumn{1}{c}{(K)} & 
\multicolumn{1}{c}{(K)} & 
\multicolumn{1}{c}{(cm)} & 
\multicolumn{1}{c}{($L_\odot$)} &  
\multicolumn{1}{c}{($M_\odot$)} &
\multicolumn{1}{c}{(years)}
\\   
\hline  
\hline
\\
M\,$33-1 $   & & & $ 46 $ & $ 8.46 $ & $ 708 $ & $ 6749 $  & $ 16.04 $ & $ 5.60 $  &  $0.026 $  & $  34.79  $  & & & $ 18 $ & $ 10.70$ & $ 1218$ & $ 24602 $ & $ 15.80 $ & $ 5.68 $   &  $0.064  $  &  $   19.8  $   \\ 
M\,$33-3 $   & & & $ 34 $ & $ 1.16 $ & $ 416 $ & $ 29927 $ & $ 16.94 $ & $ 5.81 $  &  $0.096 $  & $  278.5  $  & & & $ 55 $ & $ 2.63 $ & $ 649 $ & $ 29955 $ & $ 16.38 $ & $ 5.88 $   &  $0.565  $  &  $  76.21  $   \\ 
M\,$33-4 $   & & & $ 47 $ & $ 2.00 $ & $ 390 $ & $ 29977 $ & $ 16.99 $ & $ 5.76 $  &  $0.156 $  & $  307.3  $  & & & $ 75 $ & $ 3.88 $ & $ 599 $ & $ 29991 $ & $ 16.40 $ & $ 5.80 $   &  $1.182  $  &  $  80.14  $   \\ 
M\,$33-7 $   & & & $ 29 $ & $ 2.67 $ & $ 381 $ & $ 29931 $ & $ 17.06 $ & $ 5.85 $  &  $0.222 $  & $  360.5  $  & & & $ 57 $ & $ 4.79 $ & $ 599 $ & $ 29973 $ & $ 16.43 $ & $ 5.86 $   &  $2.170  $  &  $  85.98  $   \\ 
N\,$300-1 $  & & & $ 4 $  & $ 5.97 $ & $ 506 $ & $ 17129 $ & $ 16.65 $ & $ 5.74 $  &  $0.083 $  & $  141.1  $  & & & $ 8 $  & $ 8.57 $ & $ 959 $ & $ 28956 $ & $ 16.09 $ & $ 5.83 $   &  $0.744  $  &  $  39.24  $   \\ 
N\,$2403-2 $ & & & $ 30 $ & $ 0.90 $ & $ 440 $ & $ 29988 $ & $ 16.85 $ & $ 5.76 $  &  $0.076 $  & $    227  $  & & & $ 26 $ & $ 2.56 $ & $ 676 $ & $ 29942 $ & $ 16.34 $ & $ 5.84 $   &  $0.292  $  &  $  68.79  $   \\ 
N\,$2403-3 $ & & & $ 8 $  & $ 76.36$ & $ 455 $ & $ 2533 $  & $ 16.38 $ & $ 5.88 $  &  $0.263 $  & $  76.89  $  & & & $ 2 $  & $ 25.27$ & $ 499 $ & $ 4981 $  & $ 16.11 $ & $ 5.88 $   &  $2.821  $  &  $   40.8  $   \\ 
N\,$2403-4 $ & & & $ 100$ & $ 5.39 $ & $ 398 $ & $ 3790 $  & $ 16.62 $ & $ 5.86 $  &  $0.045 $  & $  131.7  $  & & & $ 146$ & $ 10.00$ & $ 568 $ & $ 4545 $  & $ 15.93 $ & $ 5.89 $   &  $0.585  $  &  $  26.78  $   \\ 
N\,$2403-5 $ & & & $ 27 $ & $ 1.97 $ & $ 429 $ & $ 16146 $ & $ 16.85 $ & $ 5.84 $  &  $0.109 $  & $  224.6  $  & & & $ 5 $  & $ 3.56 $ & $ 662 $ & $ 20750 $ & $ 16.34 $ & $ 5.98 $   &  $0.621  $  &  $  70.06  $   \\ 
M\,$81-5 $   & & & $ 1 $  & $ 7.60 $ & $ 428 $ & $ 3050 $  & $ 16.40 $ & $ 5.71 $  &  $0.117 $  & $  78.89  $  & & & $ 2 $  & $ 35.00$ & $ 1117$ & $ 29778 $ & $ 15.86 $ & $ 5.58 $   &  $0.296  $  &  $  23.14  $   \\ 
M\,$81-6 $   & & & $0.9 $ & $ 8.61 $ & $ 504 $ & $ 4897 $  & $ 16.31 $ & $ 5.55 $  &  $0.086 $  & $  64.76  $  & & & $0.5$  & $ 46.97$ & $ 985 $ & $ 13885 $ & $ 15.73 $ & $ 5.49 $   &  $0.226  $  &  $  17.15  $   \\ 
M\,$81-11 $  & & & $ 17 $ & $ 2.56 $ & $ 463 $ & $ 10825 $ & $ 16.64 $ & $ 5.68 $  &  $0.042 $  & $  138.4  $  & & & $ 3 $  & $ 4.54 $ & $ 746 $ & $ 15015 $ & $ 16.09 $ & $ 5.82 $   &  $0.307  $  &  $  38.59  $   \\ 
M\,$81-12 $  & & & $ 12 $ & $ 4.31 $ & $ 365 $ & $ 5548 $  & $ 16.84 $ & $ 5.89 $  &  $0.105 $  & $  217.5  $  & & & $ 10 $ & $ 7.84 $ & $ 529 $ & $ 7910 $  & $ 16.16 $ & $ 5.92 $   &  $1.275  $  &  $  46.34  $   \\ 
M\,$81-14 $  & & & $ 14 $ & $ 20.08$ & $ 341 $ & $ 4839 $  & $ 16.91 $ & $ 5.97 $  &  $0.591 $  & $  260.4  $  & & & $ 8 $  & $ 29.47$ & $ 416 $ & $ 4528 $  & $ 16.25 $ & $ 5.93 $   &  $8.513  $  &  $  56.65  $   \\ 
N\,$7793-3 $ & & &  \dots &   \dots  &   \dots &   \dots   &   \dots   &   \dots   &   \dots    &    \dots     & & &  \dots &   \dots  &  \dots  &   \dots   &   \dots   & \dots      &   \dots     &     \dots      \\ 
N\,$7793-9 $ & & & $ 46 $ & $ 4.61 $ & $ 432 $ & $ 14632 $ & $ 16.73 $ & $ 5.61 $  &  $0.101 $  & $  169.2  $  & & & $ 27 $ & $ 7.04 $ & $ 713 $ & $ 22072 $ & $ 16.18 $ & $ 5.70 $   &  $0.825  $  &  $  47.86  $   \\ 
N\,$7793-10 $   & & & $ 32 $ & $ 3.80 $ & $ 396 $ & $ 7406 $  & $ 16.85 $ & $5.92$ &  $0.121 $  & $  222.3  $  & & & $ 25 $ & $ 6.36 $ & $ 609 $ & $ 12175 $ & $ 16.24 $ & $5.99$     &  $1.174  $  &  $  55.18  $   \\ 
N\,$7793-13 $   & & & $ 32 $ & $ 1.90 $ & $ 369 $ & $ 29943 $ & $ 17.15 $ & $5.99$ &  $0.377 $  & $  452.2  $  & & & $ 42 $ & $ 4.01 $ & $ 546 $ & $ 29989 $ & $ 16.59 $ & $6.05$     &  $2.431  $  &  $  122.6  $   \\ 
IRC$+10420 $ & & & $ 222$ & $ 8.06 $ & $ 399 $ & $ 8157 $  & $ 16.79 $ & $ 5.76 $  &  $0.030 $  & $  194.2  $  & & & $ 240$ & $ 11.86$ & $ 835 $ & $ 11780 $ & $ 15.80 $ & $ 5.55 $   &  $1.900  $  &  $  20.17  $   \\ 
M\,$33$\,Var\,A & & & $ 43 $ & $ 2.29 $ & $ 536 $ & $ 11741 $ & $ 16.27 $ & $5.23$ &  $0.003 $  & $  58.86  $  & & & $ 8 $  & $ 4.11 $ & $ 1046$ & $ 14549 $ & $ 15.56 $ & $5.29$     &  $0.050  $  &  $  11.54  $   \\ 
$\eta$\,Car  & & & $ 490$ & $ 4.55 $ & $ 361 $ & $ 18134 $ & $ 17.41 $ & $ 6.57 $  &  $5.611 $  & $    809  $  & & & $ 853$ & $ 7.99 $ & $ 468 $ & $ 26164 $ & $ 17.02 $ & $ 6.74 $   &  $18.615 $  &  $  335.1  $   \\
\\
\hline  
\hline  
\end{tabular} 
\end{tiny} 
\end{center}  
\end{table*}
\end{landscape}

\clearpage

\begin{landscape}
\begin{table*}[p]   
\begin{center}   
\begin{tiny}  
\caption{Best Fit Models for Graphitic Dust and Fixed Temperature} 
\label{tab:graphitic}   
\begin{tabular}{rrrrrrrrrrrrrrrrrrrrrr}   
\\   
\hline  
\hline 
\multicolumn{2}{c}{} &  
\multicolumn{5}{c}{$T_* = 5000\,K$} &  
\multicolumn{1}{c}{} & 
\multicolumn{5}{c}{$T_* = 7500\,K$} &   
\multicolumn{1}{c}{} & 
\multicolumn{5}{c}{$T_* = 20000\,K$} 
\\
\multicolumn{1}{c}{ID} &  
\multicolumn{1}{c}{} &
\multicolumn{1}{c}{$\tau_V$} & 
\multicolumn{1}{c}{$T_d$} & 
\multicolumn{1}{c}{$\log\left(R_{in}\right)$} &  
\multicolumn{1}{c}{$\log L_*$} &
\multicolumn{1}{c}{$M_e$} &  
\multicolumn{1}{c}{} &
\multicolumn{1}{c}{$\tau_V$} & 
\multicolumn{1}{c}{$T_d$} & 
\multicolumn{1}{c}{$\log\left(R_{in}\right)$} &  
\multicolumn{1}{c}{$\log L_*$} &  
\multicolumn{1}{c}{$M_e$} &
\multicolumn{1}{c}{} &
\multicolumn{1}{c}{$\tau_V$} & 
\multicolumn{1}{c}{$T_d$} & 
\multicolumn{1}{c}{$\log\left(R_{in}\right)$} &   
\multicolumn{1}{c}{$\log L_*$} &
\multicolumn{1}{c}{$M_e$} 
\\ 
\multicolumn{3}{c}{} &  
\multicolumn{1}{c}{(K)} & 
\multicolumn{1}{c}{(cm)} & 
\multicolumn{1}{c}{($L_\odot$)} &
\multicolumn{1}{c}{($M_\odot$)}  &   
\multicolumn{2}{c}{} & 
\multicolumn{1}{c}{(K)} & 
\multicolumn{1}{c}{(cm)} & 
\multicolumn{1}{c}{($L_\odot$)} &
\multicolumn{1}{c}{($M_\odot$)}  &
\multicolumn{2}{c}{} & 
\multicolumn{1}{c}{(K)} & 
\multicolumn{1}{c}{(cm)} & 
\multicolumn{1}{c}{($L_\odot$)} &
\multicolumn{1}{c}{($M_\odot$)}   
\\
\hline  
\hline
\\
M\,$33-1 $   & & $ 8.17 $ & $ 595 $ & $ 16.15 $  & $ 5.58 $  &  $ 0.102   $   & & $ 8.05 $ & $ 701 $ & $ 16.09 $ & $ 5.62 $  &  $ 0.077  $   & & $ 5.88 $ & $ 900 $ & $ 15.96 $ & $ 5.66 $   &  $ 0.031   $  \\ 
M\,$33-3 $   & & $ 2.50 $ & $ 400 $ & $ 16.61 $  & $ 5.69 $  &  $ 0.261   $   & & $ 2.67 $ & $ 460 $ & $ 16.54 $ & $ 5.59 $  &  $ 0.202  $   & & $ 1.89 $ & $ 400 $ & $ 16.92 $ & $ 5.76 $   &  $ 0.821   $  \\ 
M\,$33-4 $   & & $ 3.14 $ & $ 400 $ & $ 16.56 $  & $ 5.58 $  &  $ 0.260   $   & & $ 3.52 $ & $ 400 $ & $ 16.68 $ & $ 5.58 $  &  $ 0.507  $   & & $ 4.53 $ & $ 614 $ & $ 16.26 $ & $ 5.72 $   &  $ 0.094   $  \\ 
M\,$33-7 $   & & $ 3.69 $ & $ 400 $ & $ 16.59 $  & $ 5.64 $  &  $ 0.351   $   & & $ 4.11 $ & $ 400 $ & $ 16.71 $ & $ 5.66 $  &  $ 0.679  $   & & $ 3.18 $ & $ 400 $ & $ 16.91 $ & $ 5.74 $   &  $ 1.321   $  \\ 
N\,$300-1 $  & & $ 6.52 $ & $ 501 $   & $ 16.41 $  & $5.72$  &  $ 0.271   $   & & $ 6.89 $ & $ 504 $ & $ 16.52 $ & $ 5.74 $  &  $ 0.475  $   & & $ 5.47 $ & $ 500 $ & $ 16.69 $ & $ 5.75 $   &  $ 0.824   $  \\ 
N\,$2403-2 $ & & $ 3.13 $ & $ 401 $  & $ 16.63 $  & $ 5.72$  &  $ 0.358   $   & & $ 2.99 $ & $ 493 $ & $ 16.45 $ & $ 5.56 $  &  $ 0.149  $   & & $ 1.89 $ & $ 408 $ & $ 16.88 $ & $ 5.73 $   &  $ 0.683   $  \\ 
N\,$2403-3 $ & & $ 18.58 $ & $ 404 $ & $ 16.71 $  & $ 5.86$  &  $ 3.070   $   & & $ 9.41 $ & $ 405 $ & $ 16.81 $ & $ 5.86 $  &  $ 2.466  $   & & $ 6.55 $ & $ 417 $ & $ 16.92 $ & $ 5.83 $   &  $ 2.849   $  \\ 
N\,$2403-4 $ & & $ 5.91 $ & $ 400 $  & $ 16.74 $  & $ 5.93$  &  $ 1.121   $   & & $ 7.09 $ & $ 300 $ & $ 17.32 $ & $ 6.33 $  &  $ 19.435 $   & & $ 4.92 $ & $ 1182$ & $ 15.75 $ & $ 5.91 $   &  $ 0.010   $  \\ 
N\,$2403-5 $ & & $ 2.78 $ & $ 400 $  & $ 16.71 $  & $ 5.89$  &  $ 0.460   $   & & $ 2.96 $ & $ 400 $ & $ 16.82 $ & $ 5.87 $  &  $ 0.812  $   & & $ 1.55 $ & $ 400 $ & $ 16.99 $ & $ 5.91 $   &  $ 0.928   $  \\ 
M\,$81-5 $   & & $ 14.32 $ & $ 514 $   & $ 16.32 $ & $5.58$  &  $ 0.393   $   & & $ 16.83 $ & $ 548 $ & $ 16.35 $ & $ 5.55$  &  $ 0.530  $   & & $ 19.68 $ & $ 587$ & $ 16.41 $ & $ 5.54 $   &  $ 0.817   $  \\ 
M\,$81-6 $   & & $ 14.55 $ & $ 535 $   & $ 16.25 $ & $5.52$  &  $ 0.289   $   & & $ 16.18 $ & $ 559 $ & $ 16.30 $ & $ 5.50$  &  $ 0.405  $   & & $ 13.09 $ & $ 558$ & $ 16.45 $ & $ 5.49 $   &  $ 0.653   $  \\ 
M\,$81-11 $  & & $ 2.81 $ & $ 406 $   & $ 16.66 $ & $ 5.80$  &  $ 0.369   $   & & $ 2.67 $ & $ 498 $ & $ 16.47 $ & $ 5.64 $  &  $ 0.146  $   & & $ 1.47 $ & $ 432 $ & $ 16.85 $ & $ 5.78 $   &  $ 0.463   $  \\ 
M\,$81-12 $  & & $ 3.77 $ & $ 399 $   & $ 16.67 $ & $ 5.79$  &  $ 0.518   $   & & $ 3.99 $ & $ 400 $ & $ 16.78 $ & $ 5.79 $  &  $ 0.910  $   & & $ 2.59 $ & $ 400 $ & $ 16.95 $ & $ 5.81 $   &  $ 1.295   $  \\ 
M\,$81-14 $  & & $ 21.78 $ & $ 350 $  & $ 16.91 $ & $ 5.97$  &  $ 9.040   $   & & $ 21.43 $ & $ 400$ & $ 16.81 $ & $ 5.84 $  &  $ 5.613  $   & & $ 22.28 $ & $ 430$ & $ 16.90 $ & $ 5.83 $   &  $ 8.834   $  \\ 
N\,$7793-3 $ & &   \dots   & \dots    &  \dots    &  \dots   &    \dots       & &   \dots   &  \dots &   \dots   &   \dots   &    \dots      & &   \dots   &  \dots &   \dots   &  \dots     &    \dots      \\ 
N\,$7793-9 $ & & $ 7.50 $ & $ 400 $  & $ 16.61 $  & $ 5.66 $ &  $ 0.506   $   & & $ 8.00 $ & $ 400 $ & $ 16.77 $ & $ 5.76 $  &  $ 0.629  $   & & $ 7.00 $ & $ 402 $ & $ 17.02 $ & $ 5.96 $   &  $ 1.516   $  \\ 
N\,$7793-10$ & & $ 2.73 $ & $ 500 $ & $ 16.19 $  & $ 5.30 $  &  $ 0.584   $   & & $ 2.74 $ & $ 500 $ & $ 16.27 $ & $ 5.25 $  &  $ 1.030  $   & & $ 1.12 $ & $ 500 $ & $ 16.45 $ & $ 5.30 $   &  $ 1.371   $  \\ 
N\,$7793-13$ & & $ 5.74 $ & $ 300 $ & $ 17.34 $  & $ 6.60 $  &  $ 0.528   $   & & $ 5.89 $ & $ 400 $ & $ 17.16 $ & $ 6.53 $  &  $ 0.923  $   & & $ 4.29 $ & $ 400 $ & $ 17.32 $ & $ 6.55 $   &  $ 1.224   $  \\ 
IRC$+10420 $ & & $ 5.08 $ & $ 401 $  & $ 16.60 $  & $ 5.66 $ &  $ 0.782   $   & & $ 5.25 $ & $ 424 $ & $ 16.64 $ & $ 5.63 $  &  $ 1.743  $   & & $ 4.00 $ & $ 400 $ & $ 16.89 $ & $ 5.70 $   &  $ 4.823   $  \\ 
M\,$33$\,Var\,A & & $ 3.37 $ & $400$ & $ 16.72 $  & $ 5.90 $ &  $ 0.041   $   & & $ 3.59 $ & $ 400 $ & $ 16.83 $ & $ 5.89 $  &  $ 0.060  $   & & $ 2.18 $ & $ 400 $ & $ 17.00 $ & $ 5.92 $   &  $ 0.056   $  \\ 
$\eta$\,Car & & $ 3.67 $ & $ 400 $   & $ 16.68 $  & $ 5.82 $ &  $ 17.258  $   & & $ 3.86 $ & $ 400 $ & $ 16.79 $ & $ 5.81 $  &  $ 7.732  $   & & $ 2.45 $ & $ 400 $ & $ 16.95 $ & $ 5.83 $   &  $ 11.767  $  \\
\\
\hline  
\hline  
\end{tabular} 
\end{tiny} 
\end{center}  
\end{table*}
\end{landscape}

\clearpage

\begin{landscape}
\begin{table*}[p]   
\begin{center}   
\begin{tiny}  
\caption{Best Fit Models for Silicate Dust and Fixed Temperature} 
\label{tab:silicate}  
\begin{tabular}{rrrrrrrrrrrrrrrrrrrrrr} 
\\  
\hline 
\hline  
\multicolumn{2}{c}{} &   
\multicolumn{5}{c}{$T_* = 5000\,K$} & 
\multicolumn{1}{c}{} &  
\multicolumn{5}{c}{$T_* = 7500\,K$} &   
\multicolumn{1}{c}{} &   
\multicolumn{5}{c}{$T_* = 20000\,K$} 
\\ 
\multicolumn{1}{c}{ID} &  
\multicolumn{1}{c}{} &  
\multicolumn{1}{c}{$\tau_V$} &  
\multicolumn{1}{c}{$T_d$} &   
\multicolumn{1}{c}{$\log\left(R_{in}\right)$} &
\multicolumn{1}{c}{$\log L_*$} &   
\multicolumn{1}{c}{$M_e$} & 
\multicolumn{1}{c}{} & 
\multicolumn{1}{c}{$\tau_V$} & 
\multicolumn{1}{c}{$T_d$} &  
\multicolumn{1}{c}{$\log\left(R_{in}\right)$} &
\multicolumn{1}{c}{$\log L_*$} &   
\multicolumn{1}{c}{$M_e$} & 
\multicolumn{1}{c}{} & 
\multicolumn{1}{c}{$\tau_V$} &  
\multicolumn{1}{c}{$T_d$} &  
\multicolumn{1}{c}{$\log\left(R_{in}\right)$} &   
\multicolumn{1}{c}{$\log L_*$}  &   
\multicolumn{1}{c}{$M_e$}   
\\   
\multicolumn{3}{c}{} & 
\multicolumn{1}{c}{(K)} &   
\multicolumn{1}{c}{(cm)} & 
\multicolumn{1}{c}{($L_\odot$)} & 
\multicolumn{1}{c}{($M_\odot$)}  &  
\multicolumn{2}{c}{} &  
\multicolumn{1}{c}{(K)} &   
\multicolumn{1}{c}{(cm)} & 
\multicolumn{1}{c}{($L_\odot$)} &   
\multicolumn{1}{c}{($M_\odot$)}  &   
\multicolumn{2}{c}{} & 
\multicolumn{1}{c}{(K)} & 
\multicolumn{1}{c}{(cm)} & 
\multicolumn{1}{c}{($L_\odot$)} &  
\multicolumn{1}{c}{($M_\odot$)} 
\\
\hline
\hline
\\
M\,$33-1 $  & & $ 14.42 $ & $ 816 $   & $ 15.58 $  & $5.56$  &  $0.013  $  & & $ 14.26 $ & $ 973$ & $  15.59 $ &  $ 5.59 $  &  $ 0.014 $   & & $ 10.86 $ & $1204$ & $ 15.75 $ &  $ 5.67 $  &  $ 0.022  $   \\ 
M\,$33-3 $  & & $ 4.30 $ & $ 592 $ & $ 15.83 $  &  $ 5.75 $  &  $0.012  $  & & $ 4.66 $ & $ 654 $ & $  15.88 $ &  $ 5.68 $  &  $ 0.017 $   & & $ 3.36 $ & $ 696 $ & $ 16.20 $ &  $ 5.77 $  &  $ 0.053  $   \\ 
M\,$33-4 $  & & $ 5.26 $ & $ 513 $ & $ 15.90 $  &  $ 5.65 $  &  $0.021  $  & & $ 5.64 $ & $ 601 $ & $  15.89 $ &  $ 5.58 $  &  $ 0.021 $   & & $ 2.58 $ & $ 400 $ & $ 16.87 $ &  $ 5.66 $  &  $ 0.892  $   \\ 
M\,$33-7 $  & & $ 5.91 $ & $ 523 $ & $ 15.91 $  &  $ 5.68 $  &  $0.025  $  & & $ 6.49 $ & $ 600 $ & $  15.93 $ &  $ 5.64 $  &  $ 0.030 $   & & $ 5.41 $ & $ 604 $ & $ 16.31 $ &  $ 5.79 $  &  $ 0.142  $   \\ 
N\,$300-1 $ & & $ 10.72 $ & $ 724 $  & $ 15.76 $  & $5.77 $  &  $0.022  $  & & $ 10.99 $ & $ 800$ & $  15.80 $ &  $ 5.74 $  &  $ 0.027 $   & & $ 9.17 $ & $ 922 $ & $ 16.03 $ &  $ 5.80 $  &  $ 0.066  $   \\ 
N\,$2403-2$ & & $ 5.56 $ & $ 583 $  & $ 15.89 $  &  $ 5.80 $ &  $0.021  $  & & $ 5.66 $ & $ 641 $ & $  15.92 $ &  $ 5.73 $  &  $ 0.025 $   & & $ 3.65 $ & $ 702 $ & $ 16.20 $ &  $ 5.78 $  &  $ 0.058  $   \\ 
N\,$2403-3$ & & $ 21.93 $ & $ 469 $ & $ 16.18 $  &  $ 5.90 $ &  $0.316  $  & & $ 19.50 $ & $ 501$ & $  16.24 $ &  $ 5.91 $  &  $ 0.370 $   & & $ 16.98 $ & $ 651$ & $ 16.31 $ &  $ 5.85 $  &  $ 0.445  $   \\ 
N\,$2403-4$ & & $ 9.81 $ & $ 505 $  & $ 16.09 $  &  $ 5.93 $ &  $0.093  $  & & $ 9.72 $ & $ 1114$ & $  15.58 $ &  $ 5.82 $  &  $ 0.009 $   & & $ 7.95 $ & $ 1499$ & $ 15.64 $ &  $ 5.85 $  &  $ 0.010  $   \\ 
N\,$2403-5$ & & $ 5.30 $ & $ 502 $  & $ 16.10 $  &  $ 6.03 $ &  $0.053  $  & & $ 5.35 $ & $ 600 $ & $  16.06 $ &  $ 5.93 $  &  $ 0.044 $   & & $ 3.41 $ & $ 628 $ & $ 16.39 $ &  $ 6.00 $  &  $ 0.129  $   \\ 
M\,$81-5 $  & & $ 28.38 $ & $ 616 $   & $ 15.86 $  & $5.63$  &  $0.094  $  & & $ 28.71 $ & $ 705$ & $  15.85 $ &  $ 5.59 $  &  $ 0.090 $   & & $ 24.38 $ & $ 913$ & $ 15.94 $ &  $ 5.59 $  &  $ 0.116  $   \\ 
M\,$81-6 $  & & $ 37.84 $ & $ 655 $   & $ 15.79 $  & $5.55$  &  $0.090  $  & & $ 35.40 $ & $ 722$ & $  15.81 $ &  $ 5.53 $  &  $ 0.093 $   & & $ 29.17 $ & $ 957$ & $ 15.87 $ &  $ 5.52 $  &  $ 0.101  $   \\ 
M\,$81-11 $ & & $ 5.26 $ & $ 586 $   & $ 15.93 $  &  $5.90$  &  $0.024  $  & & $ 5.38 $ & $ 650 $ & $  15.96 $ &  $ 5.83 $  &  $ 0.028 $   & & $ 3.36 $ & $ 702 $ & $ 16.25 $ &  $ 5.88 $  &  $ 0.067  $   \\ 
M\,$81-12 $ & & $ 7.16 $ & $ 410 $   & $ 16.26 $  &  $6.00$  &  $0.149  $  & & $ 7.33 $ & $ 500 $ & $  16.21 $ &  $ 5.93 $  &  $ 0.121 $   & & $ 5.62 $ & $ 517 $ & $ 16.55 $ &  $ 6.05 $  &  $ 0.444  $   \\ 
M\,$81-14 $ & & $ 26.85 $ & $ 400 $  & $ 16.34 $  &  $5.95$  &  $0.808  $  & & $ 25.47 $ & $ 496$ & $  16.23 $ &  $ 5.85 $  &  $ 0.462 $   & & $ 21.68 $ & $ 601$ & $ 16.36 $ &  $ 5.85 $  &  $ 0.715  $   \\ 
N\,$7793-3$ & &   \dots   & \dots &      \dots   &    \dots  &   \dots     & &   \dots  &  \dots  &    \dots   &    \dots  &     \dots     & &  \dots   &  \dots  &  \dots   &   \dots     &    \dots      \\ 
N\,$7793-9$ & & $ 8.42 $ & $ 585 $  & $ 15.85 $  &  $5.68 $  &  $0.027  $  & & $ 8.67 $ & $ 617 $ & $  15.93 $ &  $ 5.65 $  &  $ 0.039 $   & & $ 6.80 $ & $ 701 $ & $ 16.16 $ &  $ 5.70 $  &  $ 0.089  $   \\ 
N\,$7793-10 $ & & $ 6.24 $ & $ 472 $ & $ 16.17 $  &  $6.05$  &  $0.086  $  & & $ 6.56 $ & $ 512 $ & $  16.23 $ &  $ 6.03 $  &  $ 0.119 $   & & $ 4.56 $ & $ 600 $ & $ 16.44 $ &  $ 6.05 $  &  $ 0.217  $   \\ 
N\,$7793-13 $ & & $ 6.53 $ & $ 484 $ & $ 16.10 $  &  $5.94$  &  $0.065  $  & & $ 6.88 $ & $ 500 $ & $  16.21 $ &  $ 5.93 $  &  $ 0.114 $   & & $ 4.81 $ & $ 589 $ & $ 16.41 $ &  $ 5.95 $  &  $ 0.200  $   \\ 
IRC$+10420 $  & & $ 10.99 $ & $ 600 $ & $ 15.73 $  & $5.44$  &  $0.020  $  & & $ 11.51 $ & $ 700$ & $  15.77 $ &  $ 5.48 $  &  $ 0.025 $   & & $ 10.00 $ & $1000$ & $ 15.86 $ &  $ 5.57 $  &  $ 0.033  $   \\ 
M\,$33$\,Var\,A & & $ 5.04 $ & $793$ & $ 15.44 $  &  $5.36$  &  $0.002  $  & & $ 5.03 $ & $ 867 $ & $  15.48 $ &  $ 5.28 $  &  $ 0.003 $   & & $ 2.67 $ & $ 962 $ & $ 15.75 $ &  $ 5.33 $  &  $ 0.005  $   \\ 
$\eta$\,Car & & $ 9.94 $ & $ 250 $   & $ 17.08 $  &  $6.76$  &  $9.026  $  & & $ 10.42 $ & $ 300$ & $  17.05 $ &  $ 6.78 $  &  $ 8.245 $   & & $ 8.49 $ & $ 400 $ & $ 17.12 $ &  $ 6.80 $  &  $ 9.272  $   \\
\\
\hline  
\hline  
\end{tabular} 
\end{tiny} 
\end{center}  
\end{table*}
\end{landscape}

\end{document}